\documentclass[twocolumn,pra, amsmath,amssymb,showpacs,superscriptaddress,aps]{revtex4-1}

\usepackage{epsfig}
\usepackage{color}
\usepackage{graphicx}
\usepackage{bm}
\usepackage{epstopdf}

\usepackage{setspace}

\usepackage{epsfig}
\usepackage{dcolumn}

\usepackage{braket}
\usepackage{amsmath}

\newcommand{\abs}[1]{\left| #1 \right|}

\begin{document}

\title{Effective Approach to Impurity Dynamics in One-Dimensional Trapped Bose Gases}

\author{S.~I. \surname{Mistakidis}}
\affiliation{Zentrum f{\"u}r Optische Quantentechnologien, Universit{\"a}t Hamburg,
Luruper Chaussee 149, 22761 Hamburg, Germany}

\author{A.~G. \surname{Volosniev}}
\affiliation{Institut f{\"u}r Kernphysik, Technische Universit{\"a}t Darmstadt, 64289 Darmstadt, Germany}
\author{N.~T. \surname{Zinner}}
\affiliation{Department of Physics and Astronomy, Aarhus University, DK-8000 Aarhus C,  Denmark}
\affiliation{Aarhus Institute of Advanced Studies, Aarhus University, DK-8000 Aarhus C, Denmark}
\author{P. Schmelcher}
\affiliation{Zentrum f{\"u}r Optische Quantentechnologien, Universit{\"a}t Hamburg,
Luruper Chaussee 149, 22761 Hamburg, Germany}
\affiliation{The Hamburg Centre for Ultrafast Imaging, Universit{\"a}t Hamburg,
Luruper Chaussee 149, 22761 Hamburg, Germany}

\date{\today}

\begin{abstract}
We investigate a temporal evolution of an impurity atom in a one-dimensional trapped Bose gas
following a sudden change of the boson-impurity interaction strength. Our focus is on the effects of inhomogeneity due to the
harmonic confinement.
These effects can be described by an effective
one-body model where both the mass and the spring constant are renormalized.
This is in contrast to the classic renormalization, which addresses only the mass.
We propose an effective single-particle Hamiltonian and
apply the multi-layer multi-configuration time-dependent Hartree method
for bosons to explore its validity. 
Numerical results suggest that the effective mass is smaller than the impurity mass, 
which means that it cannot straightforwardly be extracted from translationally-invariant models.
\end{abstract}

\maketitle

\section{Introduction}
 Low-energy quantum states of an impurity
in a homogeneous infinite environment can often be
characterized by only the momentum $P$ of the impurity, such that the energies are
\begin{equation}
E(P)\simeq \epsilon + \frac{P^2}{2 m_{\mathrm{eff}}},
\label{eq:energy}
\end{equation} 
where $\epsilon$ and $m_{\mathrm{eff}}$ are parameters. $E(P)$ is the energy of a free particle with the mass $m_{\mathrm{eff}}$, which enables the notion of a polaron:  A quasiparticle that describes low-energy properties of an impurity in a medium~\cite{pekar1951, polaron2000}. The idea of a polaron was put forward in solid state physics to understand the motion of an electron in a polarizable solid~\cite{landau1948}. Nowadays, the polaron concept is used to describe other impurities such as a $^3$He particle in superfluid $^4$He~\cite{baym2008}, a proton or a $\Lambda$-baryon in nuclear matter~\cite{bishop1973, kutchera1993}, or a proton in proton conductors~\cite{braun2017}.  Concerning applications, the polaron concept is vital to computing properties of various strongly correlated electronic materials~\cite{salamon2001, lee2006}, and organic semiconductors of technological significance~\cite{gershenson2006}. 

To employ the picture of the polaron, the values of $m_{\mathrm{eff}}$ and $\epsilon$, and the limits of applicability of Eq.~(\ref{eq:energy}) must be specified. To this end, one can use experimental data or  (in the absence of such data) {\it ab initio} calculations. The latter are arguably the simplest that give insight into the interplay between one- and many-body physics. 
Still, only approximate solutions are available. Mixtures of cold atoms that put these solutions to the test~\cite{zwierlein2009, nascimbene2009, chevy2010, widera2012, grimm2012, koschorreck2012, massignan2014,arlt2016,jin2016, scazza2017, schmidt2018, tajima2018} inspire the discussion between the polaron theories and experiments, which sheds new light onto the underlying physics not only in cold atoms, but also in more practically relevant materials.

 Our paper contributes to this dialogue by studying effective descriptions of an impurity in a one-dimensional environment. The main focus is on an impurity in a harmonically-trapped Bose gas (see Refs.~\cite{wenz2013, astrakharchik2013, lindgren2014, levinsen2015} for an impurity in a trapped Fermi gas), which is a theoretical model used to analyze certain cold-atom experiments, see e.g. Ref.~\cite{catani2012}.  Recent studies have already addressed the corresponding trap effects on static~\cite{dehkharghani2015a, dehkharghani2018} and dynamic properties~\cite{sartori2013,volosniev2015, akram2016, kamenev2016, grusdt2017, lampo2017} of an impurity. The latter, however, employed various assumptions, e.g., the local density or mean-field approximations, whose validity is hard to test. Here we perform a variational beyond-mean-field analysis of the quench dynamics initiated by a sudden change of the impurity-environment interaction. Our numerical tool to explore the dynamics is the Multi-Layer Multi-Configuration Time-Dependent Hartree method for Bosons (ML-MCTDHB) \cite{cao2013, mistakidis2017, Mistakidis_cor, Katsimiga_DBs, Koutentakis_FF, expansion}. The ML-MCTDHB results allow us to identify parameters for which an effective one-body description is appropriate. They demonstrate the necessity to renormalize both the mass and the spring constant in the corresponding effective Hamiltonian, and reveal that the mass of the dressed impurity can be smaller than its bare mass.

 This work is structured as follows. Section~\ref{sec:ham} presents the Hamiltonian of interest. Section~\ref{sec:homog_case} introduces the effective Hamiltonian for an impurity in a homogeneous Bose gas. In Sec.~\ref{sec:inhomog_case} we extend the effective Hamiltonian of Sec.~\ref{sec:homog_case} to the inhomogeneous case. The effective model is explored here by using the ML-MCTDHB results for the quench dynamics. In Sec.~\ref{sec:summ} we summarize our findings and provide an outlook for future research. In Appendix~\ref{app:a} we derive a non-linear Schr{\"o}dinger equation, which is used to formulate the effective description of an impurity in a homogeneous Bose gas. Appendix~\ref{app:b} addresses the impurity's wavefunction within our effective Hamiltonian model. In Appendix~\ref{app:c} we describe the ML-MCTDHB approach. Appendix~\ref{app:d} provides details of our numerical simulations and analyses the accuracy of the ML-MCTDHB results.
Finally, Appendix~\ref{app:e} presents the dynamics of the probability density of the impurity and compares the ML-MCTDHB results with those of the effective model.

\section{Hamiltonian}
\label{sec:ham}

We consider an impurity atom in a weakly-interacting trapped one-dimensional Bose gas. 
The Hamiltonian is motivated by the cold-atom experiments~\cite{kohl2009,catani2012, kuhr2013,knap2017} 
\begin{equation}
H=\sum_{i=1}^N h(x_i)+h(y)+g\sum_{i>j}\delta(x_i-x_j)+c\sum_{i=1}^N \delta(x_i-y),
\label{eq:hamil_full}
\end{equation}
where  $x_i$ ($y$) is the coordinate of the $i$th boson (impurity), $N$ is the number of bosons.  The
short-range interatomic potentials are parametrized by $g>0$ (boson-boson) and $c>0$ (boson-impurity). The one-body Hamiltonian reads $h(x)=-\frac{\hbar^2}{2m}\frac{\partial^2}{\partial x^2}+\frac{m\Omega^2 x^2}{2}$, where $m$ is the mass and $\Omega$ is the trap frequency. The corresponding spring constant is $k\equiv m\Omega^2$. For the sake of argument, the impurity and a boson are described by the same one-body Hamiltonian~$h$. 

To find the low-energy eigenstates of $H$, one has to address an $(N+1)$-body problem. In the homogeneous case qualitative features of these states are well understood~\cite{parisi2017, grusdt2017, volosniev2017,pastukhov2017,robinson2016, campbell2017,astra2004, sacha2006, bruderer2008,kain2016}. The trapped case requires, however, further exploration.  In particular, the quench dynamics is of immediate interest because it potentially can be used to study $m_{\mathrm{eff}}$~\cite{catani2012,grusdt2017}. The analysis of the time evolution upon the change: $c=0$ at $t<0$ to $c>0$ at $t\geq 0$, which can be realized, e.g., by using Feshbach resonances~\cite{olshanii1998, chin2010}, is the main objective of our work. However, to set the stage we first consider the homogeneous case, i.e., $\Omega=0$. 

\section{Homogeneous Case}
\label{sec:homog_case}

\begin{figure}
\centerline{\includegraphics[scale=0.7]{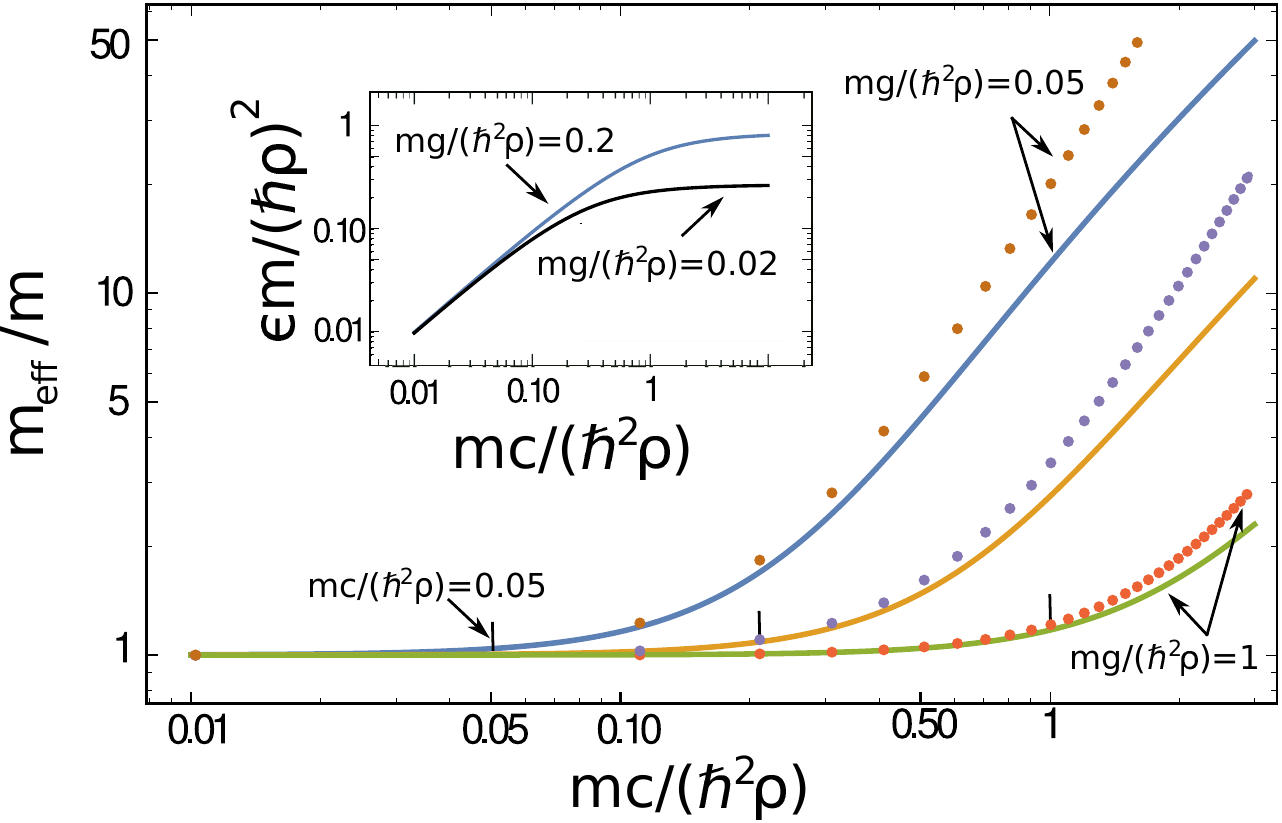}}
\caption{ The effective mass, $m_{\mathrm{eff}}/m$, as a function of the boson-impurity interaction strength, $mc/(\hbar^2\rho)$. The curves are calculated using a non-linear Schr{\"o}dinger equation derived for the ground state in Ref.~\cite{volosniev2017} and extended here to low-lying excited states, see Appendix~\ref{app:a}.  The boson-boson interaction strengths are $mg/(\hbar^2\rho)=0.05,0.2,1$ (top to bottom). The dots are the results of perturbation theory~\cite{parisi2017}, see Eq.~(\ref{eq:meff}). The short vertical lines denote $c=g$ for the corresponding curves. The inset shows the self-energy, $\epsilon m/(\hbar^2\rho^2)$, for $mg/(\hbar^2\rho)=0.02$ and $0.2$ (bottom and top) (cf.~Ref.~\cite{volosniev2017}).}
\label{fig:Fig0}
\end{figure}

 In the homogeneous case (for $N\to\infty$ and finite density  of bosons  $\rho$) the eigenstates of $H$ related to a `slow' motion of the impurity are approximately parametrized by the momentum of the impurity, see Appendix~\ref{app:a}. The corresponding energies are given by Eq.~(\ref{eq:energy}), which motivates the effective Hamiltonian 
\begin{equation}
H_{\mathrm{eff}}=\epsilon-\frac{\hbar^2}{2m_{\mathrm{eff}}}\frac{\partial^2}{\partial y^2},
\label{eq:eff_k=0}
\end{equation}
where $\epsilon(g/\rho,c/\rho)$ and $m_{\mathrm{eff}}(g/\rho,c/\rho)$ are referred to as the self-energy and the effective mass, respectively. These quantities can be measured in cold Bose gases.  For example, $\epsilon$ can be related to the frequency shift of the spectroscopic signal (cf.~\cite{arlt2016}), whereas $m_{\mathrm{eff}}$ can be extracted from the dynamics (cf.~\cite{catani2012}). The values of $\epsilon$ and $m_{\mathrm{eff}}$ can also be computed~\cite{parisi2017,grusdt2017, volosniev2017}.  We summarize the theoretical expectations on $\epsilon$ and $m_{\mathrm{eff}}$ in Fig.~\ref{fig:Fig0}. To obtain the data shown in this figure, we derive a non-linear Schr{\"o}dinger equation that describes the deformation of the Bose gas in the vicinity of the impurity (see Appendix~\ref{app:a}) for $P\neq 0$, thus, extending the theory developed in~\cite{volosniev2017} for the ground state. 
 The derived equation assumes that there is no depletion of the Bose gas and no boson-boson correlations, allowing us to solve it analytically~\cite{kamenev2016, tsuzuki1971, ishikawa1980, hakim1997}. The corresponding low-energy spectrum has the form of Eq.~(\ref{eq:energy}), providing us with $\epsilon$ and $m_{\mathrm{eff}}$. Both the self-energy and the effective mass in Fig.~\ref{fig:Fig0} are increasing functions of the boson-impurity interaction $c$: The impurity gains `weight'  when embedded in the medium ($m_{\mathrm{eff}}>m$). 
This gain can be large even in the weakly-interacting regime ($c\to 0, g\to 0$) provided that $c\gg g$; the energy correction in this regime is not important -- it scales as $\sqrt{g}$~\cite{volosniev2017}. 
Finally, we note that in the weakly-interacting regime the effective mass was derived
from perturbative calculations within the Bogoliubov approximation~\cite{parisi2017}
\begin{equation}
\frac{m_{\mathrm{eff}}}{m}=1+\frac{2c^2}{3g^{3/2}\pi}\sqrt{\frac{m}{\hbar^2\rho}}.
\label{eq:meff}
\end{equation}
For the considered cases (see~Fig.~\ref{fig:Fig0}) this relation is accurate for $c\lesssim g$.

%%%%%%%%%%%%%%%%%%%%%%%%%%%%%%%%%%%%%%%%%%%%%%%%%%%%%%%%%%%%%%%%%%%%%%%%%%%%%%%%%%%%%%%%%%%%%%%%%%%%%%%%%%%%%%%%%%%%%%%
%%%%%%%%%%%%%%%%%%%%%%%%%%%%%%%%%%%%%%%%%%%%%%%%%%%%%%%%%%%%%%%%%%%%%%%%%%%%%%%%%%%%%%%%%%%%%%%%%%%%%%%%%%%%%%%%%%%%%%%

\section{Harmonic Trap Impact}
\label{sec:inhomog_case}

 Cold-atom experiments are often performed in harmonic traps, hence, the Hamiltonian~(\ref{eq:eff_k=0}) must be modified to take this into account.  If the trap does not vary appreciably on the length scale given by the healing length, then a natural extension of Eq.~(\ref{eq:eff_k=0}) to the trapped case is 
\begin{equation}
H^{\mathrm{trap}}_{\mathrm{eff}}=\epsilon(y)-\frac{\hbar^2}{2m_{\mathrm{eff}}(y)}\frac{\partial^2}{\partial y^2}+\frac{k y^2}{2},
\label{eq:eff_total}
\end{equation}
where $\epsilon(y)\equiv \epsilon(g/\rho(y), c/\rho(y))$ and $m_{\mathrm{eff}}(y)\equiv m_{\mathrm{eff}}(g/\rho(y), c/\rho(y))$ with $\epsilon$ and $m_{\mathrm{eff}}$ from Eq.~(\ref{eq:eff_k=0}).The functions $\epsilon(y)$ and $m_{\mathrm{eff}}(y)$ are defined only if the impurity is inside the cloud, i.e., it does not probe the region with $\rho(y)=0$ where $g/\rho(y)$ and $c/\rho(y)$ cannot be specified.  To gain some insight into the properties of the Hamiltonian~(\ref{eq:eff_total}), let us consider $g,c\to 0$ with $g\gg c$. In this limit the leading order correction to the energy can be assessed using first order perturbation theory: $\epsilon \simeq c\rho$ (cf.~\cite{volosniev2017} and Fig.~\ref{fig:Fig0}). The correction to $m_{\mathrm{eff}}$ is given by Eq.~(\ref{eq:meff}), thus, $m_{\mathrm{eff}}/m-1\sim c^2$ and can be neglected in the leading order. The density profile $\rho(y)$ can be estimated from the Thomas-Fermi approximation~\cite{pethick}: $\rho_{\mathrm{TF}}(|y|<R)=\frac{kR^2}{2g}\left(1-\frac{y^2}{R^2}\right)$, and $\rho_{\mathrm{TF}}(|y|>R)=0$,  where $R=(3gN/(2 k))^{1/3}$. Therefore, the effective Hamiltonian reads
\begin{equation}
H^{\mathrm{trap}}_{\mathrm{eff}}\simeq -\frac{\hbar^2}{2m}\frac{\partial^2}{\partial y^2}+\left(1-\frac{c}{g}\right)\frac{m \Omega^2 y^2}{2}.
\label{eq:eff_k0}
\end{equation}
This equation is expected to describe qualitatively the $c<g$ regime for weak and moderate interactions, 
where the approximations used to derive Eq.~(\ref{eq:eff_k0}) are accurate. 
Three further comments are in order here: {\it i)} the spring constant has to be {\it always} renormalized (cf.~\cite{sartori2013} for highly-imbalanced Bose-Bose mixtures); {\it ii)}  for $c<g$ and small $g$ the dynamics is determined mainly by the renormalization of the spring constant; {\it iii)} for $c>g$ the harmonic oscillator in Eq.~(\ref{eq:eff_k0}) acquires a negative spring constant, hence there is no reasonable ground state of the Hamiltonian~(\ref{eq:eff_k0}).
Due to {\it iii)} below we consider the cases with $c<g$ and $c>g$ separately. 
Note that the value $c=g$ is not exceptional for the homogeneous case (see~Fig.~\ref{fig:Fig0}), hence,
the value $c=g$ is special only due to the external trap. The two regimes ($c<g$ and $c>g$) should not be confused with the miscible-immiscible phase 
separation of two harmonically trapped gases~\cite{timmermans1998, malomed2000}, which requires a finite number of interacting impurity atoms (cf.~Ref.~\cite{sartori2013}).

To renormalize the mass and spring constant beyond~Eq.~(\ref{eq:eff_k0}), maintaining 
a simple form, we propose the following Hamiltonian for $c<g$
\begin{equation}
\overline H^{\mathrm{trap}}_{\mathrm{eff}}\simeq \overline \epsilon-\frac{\hbar^2}{2\overline m_{\mathrm{eff}}}\frac{\partial^2}{\partial y^2}+\frac{\overline k_{\mathrm{eff}} y^2}{2},
\label{eq:eff_k}
\end{equation}
where $\overline m_{\mathrm{eff}}$ and $\overline k_{\mathrm{eff}}$ are independent from each another. 
The advantage of $\overline H^{\mathrm{trap}}_{\mathrm{eff}}$ 
is that it requires no knowledge of the density of bosons $\rho(y)$ -- the effective parameters 
$\overline \epsilon$, $\overline m_{\mathrm{eff}}$ and $\overline k_{\mathrm{eff}}$ 
do not depend on position. The disadvantage is that the effective parameters are not simply connected to 
the homogeneous model (see below).
Note also that even though the Hamiltonian~(\ref{eq:eff_k}) seems more complicated than the model in Eq.~(\ref{eq:eff_k=0}), the corresponding propagator 
is well-known~\cite{feynman1965, Barone2003}, which grants an easy access to various observables, see Appendix~\ref{app:b} for an example.

\subsection{Case with $c<g$} 

To test the proposed effective Hamiltonian~(\ref{eq:eff_k}),  we investigate the time evolution of the impurity
after a sudden change of $c$.
We assume that at $t<0$ the impurity is non-interacting, i.e., $c=0$, and the system is in the ground state of $H$.
At $t=0$ there is a change to $c>0$, which for $t>0$ leads to a breathing-type oscillation of the density profile of the impurity.
We study these oscillations, assuming the values
$N=100$, $\Omega=2\pi \times 20$Hz, $g=10^{-37} \mathrm{J m}$ and $m=m(^{87}\mathrm{Rb})$, which
are typical for Rb atoms in quasi-one-dimensional geometries~\cite{catani2012, egorov2013}. 
The corresponding parameters that define the Bose density are $R\simeq 19\mu$m and $\rho(0)\simeq 4/(\mu\mathrm{m})$. The dimensionless interaction strength $gm/(\rho(0)\hbar^2)$ 
is approximately $0.33$. 
 We expect the local density approximation, and hence Hamiltonian~(\ref{eq:eff_total}), 
to be accurate for these parameters.
 Indeed, the size of the hole created by an impurity in a homogeneous Bose gas
is of the order of the healing length~\cite{parisi2017,volosniev2017}
$\sqrt{\hbar^2/(gm\rho(0))}\simeq 0.4\mu$m, which
is much smaller than the relevant length scales of the problem, such as the harmonic oscillator length $\sqrt{\hbar/(m\Omega)}\simeq 2.5 \mu$m.
Note also that for these parameters and $c=g$, 
the effective mass at the center of the trap is $m_{\mathrm{eff}}(0)/m\simeq 1.1$.
The renormalization is larger at the edges of the Bose gas where
the density of bosons is smaller (see~Eq.~(\ref{eq:meff})). 
Therefore, a noticeable renormalization of $\overline m_{\mathrm{eff}}$ can be anticipated.

%%%%%%%%%%%%%%%%%%%%%%%%%%%%%%%%%%%%%%%%%%%%%%%%%%%%%%%%%%%%%%%%%%%%%%%%%%%%%%%%%%%%%%%%%%%%%%%%%%%%%%%%%%%%%%%%%%%%%%%
\begin{figure}
\centerline{\includegraphics[scale=0.35]{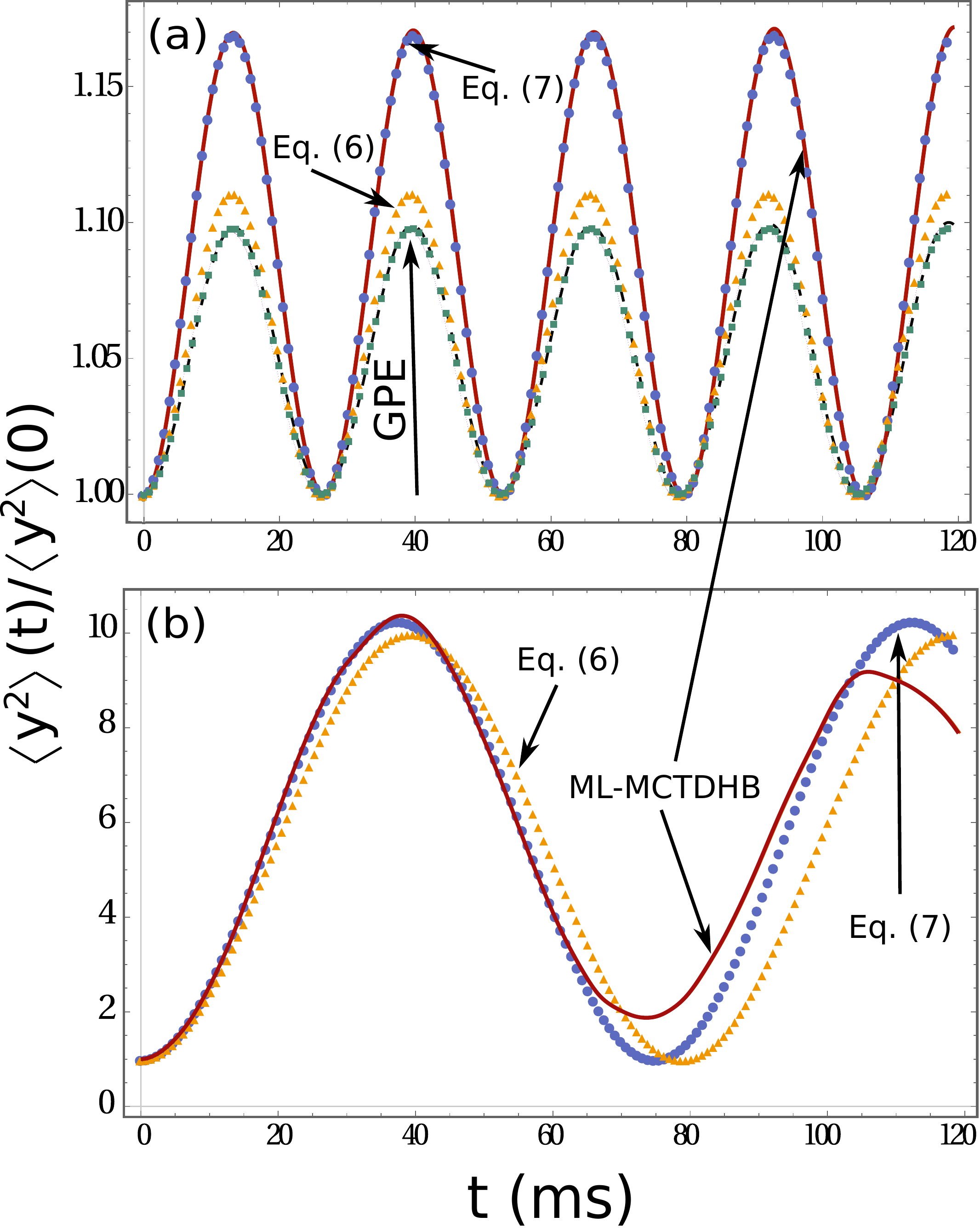}}
\caption{ The size of the impurity cloud $\langle y^2 \rangle(t)/\langle y^2 \rangle(0)$ as a function of time (in $m$s). The (red) solid curves show the ML-MCTDHB results. The blue dots illustrate the fits with the effective Hamiltonian~(\ref{eq:eff_k}). The yellow triangles are the results of the Hamiltonian~(\ref{eq:eff_k0}) (no fit parameters). The (black) dashed curve (in {\bf (a)}) shows  the result obtained using the two coupled Gross-Pitaevskii equations (GPE), the green squares are the fits with~(\ref{eq:eff_k}). Panel {\bf (a)} is for $c=0.1g$, the parameters fitted to the ML-MCTDHB results are $\overline m_{\mathrm{eff}}/m=0.983$, $\overline k_{\mathrm{eff}}/k=0.87$; to the mean-field results they are $\overline m_{\mathrm{eff}}/m=1.006$, $\overline k_{\mathrm{eff}}/k=0.9054$. Panel {\bf (b)} is for $c=0.9g$, the fit parameters are $\overline m_{\mathrm{eff}}/m=0.937$, $\overline k_{\mathrm{eff}}/k=0.104$.}
\label{fig:Fig1}
\end{figure}
%%%%%%%%%%%%%%%%%%%%%%%%%%%%%%%%%%%%%%%%%%%%%%%%%%%%%%%%%%%%%%%%%%%%%%%%%%%%%%%%%%%%%%%%%%%%%%%%%%%%%%%%%%%%%%%%%%%%%%%

We determine the size (variance) of the impurity cloud $\langle y^2\rangle$ first using the ML-MCTDHB.
In this method the many-body wavefunction is variationally optimized in a time-dependent basis.
The size of the basis is determined by looking at the convergence of the observable of interest. 
For a more detailed description of this method and estimates of its accuracy we refer to Appendices~\ref{app:c} and~\ref{app:d}, correspondingly
Then we calculate $\langle y^2\rangle$ within our one-body effective model~(\ref{eq:eff_k}). The latter model has the parameters $\overline m_{\mathrm{eff}}$ and $\overline k_{\mathrm{eff}}$, which are found by fitting to the numerical results;
see Fig.~\ref{fig:Fig1}. The parameter $\overline \epsilon$ defines an overall energy shift, which cannot be obtained from the dynamics.
We compare $\langle y^2\rangle$  for $c=0.1 g$ (weakly interacting impurity) and $c=0.9 g$ (interactions are of the same order), see Fig.~\ref{fig:Fig1}. The Hamiltonian~(\ref{eq:eff_k}) describes both cases well. For $c=0.1 g$ the agreement is excellent; for $c=0.9 g$ the model~(\ref{eq:eff_k}) is less accurate:   For long evolution times we observe an attenuation of the oscillation amplitude.  This attenuation occurs because the impurity probes the free space outside the bosonic ensemble, which is beyond the description provided by Eq.~(\ref{eq:eff_k}). Furthermore, by analogy to systems with heavy impurities~\cite{hakim1997, astra2004, sykes2009, lausch2018},
there is an energy exchange between the Bose gas and the impurity which is not included in Eq.~(\ref{eq:eff_k}) (cf.~\cite{peotta2013}).
The energy exchange is expected to increase when the impurity interacts with the tail of the Bose cloud where the parameter $c/\rho(y)$ is large, see also Ref.~\cite{mistakidis_20190} for a more detailed analysis of the energy exchange.

\begin{figure}
\centerline{\includegraphics[scale=0.55]{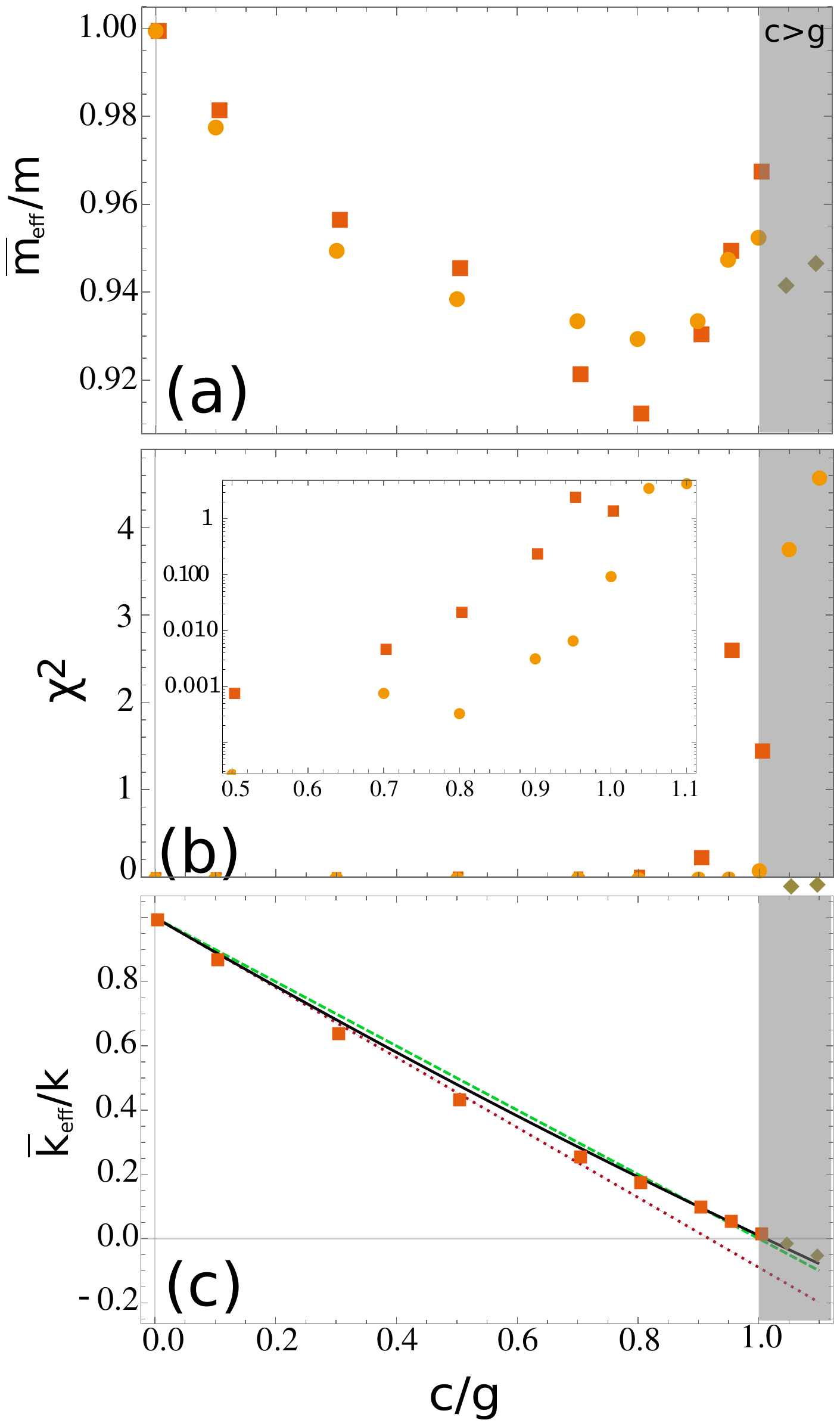}}
\caption{{\bf (a)}: The effective mass $\overline m_{\mathrm{eff}}/m$: The squares (dots) show the effective masses that minimize $\chi^2$ for the time interval $[0,t_{M_1}]$ ($[0,t_{M_2}]$). The rhombi in the $c>g$ region are obtained using the time interval $[0,32m\mathrm{s}]$.  {\bf (b)}: $\chi^2$ as a function of $c/g$. The plot markers are as in {\bf (a)}.   The inset is a semi-log plot of these data. {\bf (c)}: The effective spring constant $\overline k_{\mathrm{eff}}/k$ as a function of $c/g$. The (green) dashed line presents the mean-field result, i.e., $\overline k_{\mathrm{eff}}/k=1-c/g$, see Eq.~(\ref{eq:eff_k0}).
The solid curve shows the renormalization of the spring constant from Eq.~(\ref{eq:k_analytical}), and the (red) dotted line shows the corresponding leading order in $c$. 
Note that the results of the two fitting procedures (with $t_{M_1}$ and $t_{M_2}$) are indistinguishable on this scale.}
\label{fig:Fig2}
\end{figure}

The renormalization of the mass and spring constant has to be performed even for tiny interactions to accurately describe the ML-MCTDHB data (see also Fig.~\ref{fig:Fig2}). To fit 
the data, we must have $\overline m_{\mathrm{eff}}/m<1$, which is at odds with our intuition from the homogeneous model (see~Fig.~\ref{fig:Fig0}). 
The beyond-mean-field corrections captured by our method (cf.~\cite{kasevich2016, marchukov2017}) are crucial for this effect. Figure~\ref{fig:Fig1}{\bf (a)} presents the corresponding calculations using the two coupled Gross-Pitaevskii equations~\cite{pethick, stringari, emergent} (see also Appendix~\ref{app:c}), which disagree with the ML-MCTDHB results and suggest $\overline m_{\mathrm{eff}}/m>1$. 
The renormalization of the frequency agrees with the results of Ref.~\cite{sartori2013} derived for an imbalanced Bose-Bose mixture. However, since we work with a single impurity,
there are no collective excitations, which sets our findings apart from those in~\cite{sartori2013} even at the mean-field level.

To calculate the optimal values of $\overline m_{\mathrm{eff}}/m$ and $\overline k_{\mathrm{eff}}/k$ for $c/g\in[0.1,1]$, we minimize the sum
\begin{equation} 
\chi^2=\sum_{i=1}^M\frac{(\langle y^2 \rangle^{H}(t_i)-\langle y^2 \rangle^{\overline H^{\mathrm{trap}}_{\mathrm{eff}}}(t_i))^2}{M \langle y^2 \rangle(0)},
\label{eq:chi2}
\end{equation}
where $M$ is the number of used data points. Note that $\overline m_{\mathrm{eff}}$ and  $\overline k_{\mathrm{eff}}$
depend on the choice of the time interval $[t_1=0,t_M]$. We use two values of $t_M$ to illustrate this 
statement: $t_{M_1}=120m$s, which is the timespan of our numerical simulations  (see Fig.~\ref{fig:Fig1}), and
$t_{M_2}$, which is the time when $\langle y^2\rangle$ reaches its maximum for the first time (e.g., $t_{M_2}\simeq 13 m$s in  Fig.~\ref{fig:Fig1} {\bf (a)}).
The corresponding values for $\overline m_{\mathrm{eff}}$ and  $\overline k_{\mathrm{eff}}$ are presented in Fig.~\ref{fig:Fig2}.
The two sets agree, their difference being less than $1.5\%$ does not affect our main findings. 
Note that the values of $\overline m_{\mathrm{eff}}$ and $\overline k_{\mathrm{eff}}$ that minimize Eq.~(\ref{eq:chi2})
describe accurately also other quantities, such as overlaps and densities, see Appendix~\ref{app:e}. 
Therefore, the effective Hamiltonian $\overline H^{{\mathrm{trap}}}_{\mathrm{eff}}$ is universal, in the sense that it does not depend 
on the low-energy observable of interest.

%%%%%%%%%%%%%%%%%%%%%%%%%%%%%%%%%%%%%%%%%%%%%%%%%%%%%%%%%%%%%%%%%%%%%%%%%%%%%%%%%%%%%%%%%%%%%%%%%%%%%%%%%%%%%%%%%%%%%%%
%%%%%%%%%%%%%%%%%%%%%%%%%%%%%%%%%%%%%%%%%%%%%%%%%%%%%%%%%%%%%%%%%%%%%%%%%%%%%%%%%%%%%%%%%%%%%%%%%%%%%%%%%%%%%%%%%%%%%%%

\subsection{Case with $c>g$}  

The one-body description~(\ref{eq:eff_k}) is valid for $c/g\lesssim 1$ for all considered time intervals, which is quantified by small values of $\chi^2$.
For larger values of $c/g$ the impurity probes the edges of the cloud which is not captured by $\overline H^{{\mathrm{trap}}}_{\mathrm{eff}}$.
Still, the effective Hamiltonian describes well also the dynamics for $c>g$ at short time scales. To illustrate this, we minimize~Eq.~(\ref{eq:chi2}) using the time interval $[0,32m\mathrm{s}]$ (see Fig.~\ref{fig:Fig2}), for which the impurity is always inside the bosonic ensemble. For larger times, the impurity is pushed to the edge of the Bose gas, and the effective 
description (\ref{eq:eff_total}) becomes invalid: 
The part of the space that is not occupied by the Bose gas must be included into the effective description~\cite{footnote1}. One can extend~Eq.~(\ref{eq:eff_total})
to $y>|R|$ by using a free oscillator or some alternative effective Hamiltonian, e.g., from Ref.~\cite{dehkharghani2015a}. 
However, such an investigation goes beyond the polaron description of Eq.~(\ref{eq:eff_total}) since the density of the Bose gas vanishes for $y>|R|$.

\subsection{Beyond-mean-field corrections}   
\begin{figure}
\centerline{\includegraphics[scale=0.6]{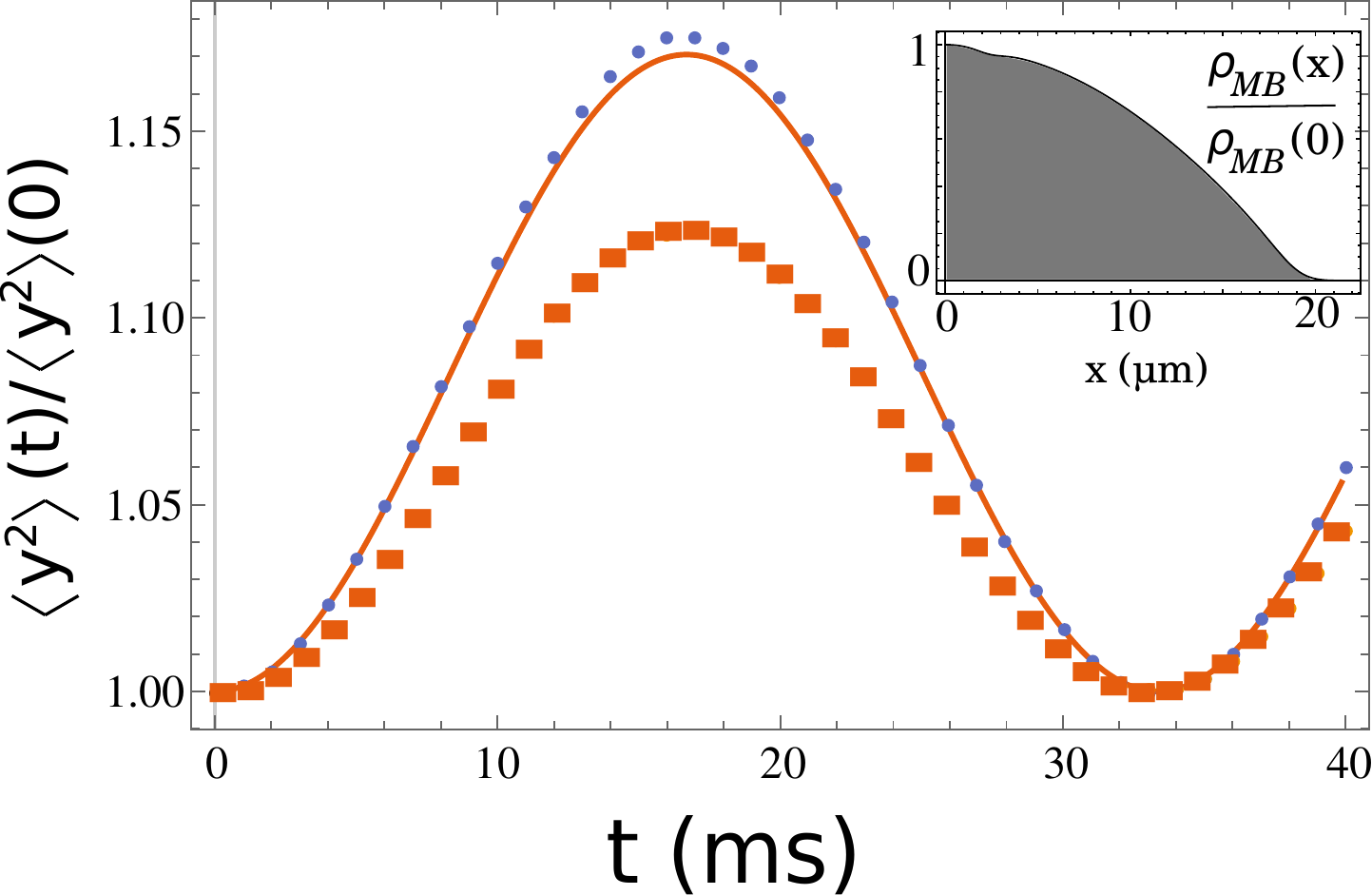}}
\caption{ The size of the impurity cloud $\langle y^2 \rangle(t)/\langle y^2 \rangle(0)$ for $c/g=0.1$ as a function of time (in $m$s),
the parameters are as in Fig.~\ref{fig:Fig1}{\bf(a)}. 
The solid (red) curve shows the ML-MCTDHB results. The blue dots are generated using Eq.~(\ref{eq:eff_total_MB}), 
the orange rectangles are obtained using $\epsilon_{BMF}(y)$ instead of $c\rho_{MB}$ in Eq.~(\ref{eq:eff_total_MB}). 
The inset shows the ML-MCTDHB density of bosons.} 
\label{fig:fig4}
\end{figure}

The results presented in Fig.~\ref{fig:Fig2} differ considerably from the mean-field 
predictions even for~$c\to~0$.  This deviation is due to the correction 
to the Thomas-Fermi density of bosons resulting from beyond-mean-field correlations, which are included in the ML-MCTDHB results. 
To show this, we determine the density of 
bosons via ML-MCTDHB, $\rho_{MB}(y)$, at $t=0$. The correspinding self-energy 
for $c\to 0$ is $\epsilon_{MB}(y)\simeq c\rho_{MB}(y)$. In this limit one has to use the Hamiltonian
\begin{equation}
H^{\mathrm{trap}}_{\mathrm{eff}}\simeq c\rho_{MB}(y)-\frac{\hbar^2}{2m}\frac{\partial^2}{\partial y^2}+\frac{k y^2}{2},
\label{eq:eff_total_MB}
\end{equation}
instead of Eq.~(\ref{eq:eff_k0}). We use the Hamiltonian~(\ref{eq:eff_total_MB})  
to calculate $\langle y^2 \rangle(t)/\langle y^2 \rangle(0)$ for the parameters of Fig.~\ref{fig:Fig1}{\bf(a)}, see~Fig.~\ref{fig:fig4}. 
The corresponding density $\rho_{MB}$ is shown in the inset of Fig.~\ref{fig:fig4}. The figure demonstrates that the Hamiltonian~(\ref{eq:eff_total_MB}) describes well the ML-MCTDHB results. Therefore, the corrections to the Thomas-Fermi profile can indeed renormalize
$\overline k_{\mathrm{eff}}$ as well as $\overline m_{\mathrm{eff}}$. 
To illustrate how $\rho_{MB}$ renormalizes the effective mass and the spring constant we express it as $\rho_{MB}(y)=\alpha^{(2)} y^2+(\rho_{MB}-\alpha^{(2)} y^2)$, where $\alpha^{(2)}$ defines the renormalization of the spring constant. For simplicity, we assume that $\alpha^{(2)}=-k/2g$, i.e., as in the Thomas-Fermi profile. The effective mass parameter mimics the action of $(\rho_{MB}(y)-\alpha^{(2)} y^2)$. For the sake of discussion, we assume that the effective mass is used to account for the transition from the ground to the second excited state of the harmonic oscillator (which is the major transition for the dynamics we observe), i.e.,
\begin{equation}
\langle 0|\frac{\hbar^2}{2\overline m_{\mathrm{eff}}}\frac{\partial^2}{\partial y^2}|2\rangle= 
\langle 0|\frac{\hbar^2}{2m}\frac{\partial^2}{\partial y^2}-c(\rho_{MB}(y)-\alpha^{(2)} y^2)|2\rangle.
\end{equation}
In the limit $c\to 0$ the effective mass reads
\begin{equation}
\frac{\overline m_{\mathrm{eff}}}{m}=1+c\frac{\langle 0|(\rho_{MB}(y)-\alpha^{(2)} y^2)|2\rangle}{\langle 0|\frac{\hbar^2}{2 m}\frac{\partial^2}{\partial y^2}|2\rangle}.
\end{equation}
For the system under consideration, this expression leads to $\frac{\overline m_{\mathrm{eff}}}{m}\simeq 1-0.594c/g$. It predicts that $\frac{\overline m_{\mathrm{eff}}}{m}<1$ for the ML-MCTDHB density profiles, in agreement with the results presented in Fig.~\ref{fig:Fig2}{\bf{(a)}}.

To provide further analytical insight we derive the following beyond-mean-field self-energy
\begin{equation}
\epsilon_{BMF}(y)\simeq c\rho_{\mathrm{TF}}+c\sqrt{\frac{gm\rho_{TF}(y)}{\hbar^2\pi^2}}-c^2\sqrt{\frac{m\rho_{TF}(y)}{8g\hbar^2}}.
\label{eq:epsilon_y}
\end{equation}
The first term here is the already discussed mean-field result. The second term is due to the beyond-Gross-Pitaevskii correction to the Thomas-Fermi density.
To calculate it, we combined the local density approximation with the fact 
that the chemical potential must be constant in the external field~\cite{landau5, footnote_density}:
$kx^2/2+{\partial \varepsilon(x)}/{\partial N}=\mathrm{const}$ (cf.~\cite{kim2003}), where $\varepsilon(x)$ is the energy of the ground state of $H$ for $c=0$, $k=0$ and a fixed density of bosons $\rho(x)$. 
The correction to $\rho_{\mathrm{TF}}$ is obtained utilizing $\varepsilon\simeq Ng\rho/2-2g^{3/2}N\sqrt{\rho}/(3\pi)\sqrt{m/\hbar^2}$~\cite{lieb1963}.
This procedure for calculating the density was successfully tested with the density-matrix renormalization group~\cite{schmidt2007}.
The last term in Eq.~(\ref{eq:epsilon_y}) is the $c^2$-term in the expansion of $\epsilon(y)$ in powers of $c$ (assuming $\rho=\rho_{\mathrm{TF}}$)~\cite{footnote_Bog}.
The energy $\epsilon_{BMF}(y)$ allows us to calculate
$\overline k_{\mathrm{eff}}/k$ for $c\to 0$ 
\begin{equation}
\frac{\overline k_{\mathrm{eff}}}{k}\simeq 1-\frac{c}{g}-\frac{c}{\sqrt{2}\hbar \Omega R \pi}+\frac{c^2}{4\hbar\Omega R g}.
\label{eq:k_analytical}
\end{equation}
This equation agrees qualitatively with the results based on ML-MCTDHB, see Fig.~\ref{fig:Fig2}{\bf{(c)}}.
It is worthwhile noting, that the beyond-mean-field correction: $i)$ vanishes in the limit 
$R, N \to \infty$ for a fixed value of $\Omega$, and $ii)$ grows in the limit $\Omega\to 0$ with fixed $N$. 
The density of bosons goes to zero in the latter case, which, indeed, implies strong beyond-mean-field correlations, in particular, 
fermionization~\cite{girardeau1960}.  We also verified $ii)$ numerically 
using $\Omega=2\pi \times 4$Hz for $c/g=0.1$ (all other parameters as before). 
We observed that the corresponding effective parameters ($\overline m_{\mathrm{eff}}/m=0.967, \overline k_{\mathrm{eff}}/k=0.826$) deviate 
stronger from the mean-field prediction than those in Fig.~\ref{fig:Fig2} for $c/g=0.1$.

 Note that the beyond-mean-field density profile of bosons $\rho_{\mathrm{TF}}+\sqrt{gm\rho_{TF}}/(\hbar \pi)$,
which produces the two first terms of $\epsilon_{BMF}$ in Eq.~(\ref{eq:epsilon_y}), is somewhat different from $\rho_{MB}$  and leads to different values of $\langle y^2 \rangle(t)/\langle y^2 \rangle(0)$, see~Fig.~\ref{fig:fig4}.
This difference is not surprising, since it is clear that Eq.~(\ref{eq:epsilon_y}) does not include all possible
corrections to the Thomas-Fermi profile. For example, Eq.~(\ref{eq:epsilon_y}) has been derived using the local density approximation,
and hence boundary effects are not fully captured (cf.~\cite{fetter1998}). The impurity dynamics is sensitive to these corrections to $\rho_{TF}$
(see~Fig.~\ref{fig:fig4}), hence, it can potentially be employed as a witness of beyond-mean-field physics to test theoretical methods 
that describe it.

%%%%%%%%%%%%%%%%%%%%%%%%%%%%%%%%%%%%%%%%%%%%%%%%%%%%%%%%%%%%%%%%%%%%%%%%%%%%%%%%%%%%%%%%%%%%%%%%%%%%%%%%%%%%%%%%%%%%%%%
%%%%%%%%%%%%%%%%%%%%%%%%%%%%%%%%%%%%%%%%%%%%%%%%%%%%%%%%%%%%%%%%%%%%%%%%%%%%%%%%%%%%%%%%%%%%%%%%%%%%%%%%%%%%%%%%%%%%%%%

\section{Summary}
\label{sec:summ}

 We explore properties of a single impurity atom in a weakly-interacting one-dimensional Bose gas. 
First, we consider the homogeneous case where the low-lying energy states can be described using the effective mass, which is larger than the bare mass of the impurity. Then we use the concepts derived in the homogeneous case to describe the time dynamics of an impurity in a Bose gas in trapped cold-atom systems.
We propose a corresponding effective Hamiltonian to analyze this dynamics. The effective mass and the spring constant are obtained from the time evolution of the size of the impurity cloud, which is calculated using the ML-MCTDHB method. The effective mass is found to be smaller than the bare mass of the impurity, which we explain using the ML-MCTDHB density profile for bosons.

Our work sets the stage for several interesting future investigations. An intriguing prospect is to consider a quench dynamics of a Bose gas with two impurities. Such an investigation should allow one to understand the importance of impurity-impurity correlations induced by the Bose gas. Another interesting direction would be to study the emergent correlation effects in Bose-Fermi or Fermi-Fermi mixtures. 

\vspace*{1em}

{\it Note added:} The limits of applicability of the model employed to describe an impurity in a homogeneous Bose gas were briefly discussed in Refs.~\cite{smith2019, panochko2019}.

\vspace*{1em}

\begin{acknowledgments}
We thank Oleksandr Marchukov for comments on the manuscript and Martin Zwierlein for bringing~Ref.~\cite{ishikawa1980} to our attention.
A.~G.~V. and N.~T.~Z. thank Hans-Werner Hammer, Amin Dehkharghani, Oleksandr Marchukov and Joachim Brand for useful discussions concerning the Bose polaron problem.  
This work was supported by the Deutsche Forschungsgemeinschaft (DFG) 
in the framework of the SFB 925 ``Light induced dynamics and control of correlated quantum systems'' (S.~I.~M. and P.~S.);
the Humboldt Foundation, the DFG (VO 2437/1-1) and the Ingenium organization of TU Darmstadt (A. G. V.); the Danish Council for
Independent Research and the DFF Sapere Aude program (N. T. Z.).

\vspace*{1em}

S.~I. Mistakidis and A.~G. Volosniev contributed equally to this work.
\end{acknowledgments}

\begin{widetext}

\appendix
\section{A non-linear Schr{\"o}dinger equation for an impurity in a Bose gas}
\label{app:a}

Here we consider the Hamiltonian (2) from the main text without external traps, i.e., with $\Omega=0$
\begin{equation}
H=-\sum_{i}\frac{\hbar^2}{2m}\frac{\partial^2}{\partial x_i^2}-\frac{\hbar^2}{2m}\frac{\partial^2}{\partial y^2}+g\sum_{i>j}\delta(x_i-x_j)+c\sum_{i}\delta(x_i-y).
\end{equation}
For the sake of argument, we assume that the particles move in the interval  $[0,2\pi R]$ with periodic boundary conditions, i.e., in a ring of radius $R$.
The basis set for this problem can be written as $\{e^{i \frac{n_1x_1+...+n_N x_N+my}{R}}\}$, where $n_j$ and $m$ are integers.  
$L\equiv \frac{1}{R}(n_1+...+n_N+m)$ is an integral of motion, therefore, an eigenfunction of $H$ can be written 
as $\Psi=e^{i y L} \psi(z_1,...,z_N)$ with $z_i=2\pi R\theta(y-x_i)+x_i-y$, where $\theta(x)$ is the Heaviside step function, i.e., 
$\theta(x>0) = 1$ and zero otherwise. The variables $z_i$ are defined such that $0\leq z_i \leq 2\pi R$. Upon inserting this 
function into the Schr{\"o}dinger equation, $H\Psi={\cal E}\Psi$, we obtain the following equation for $\psi(0<z_i<2\pi R)$~\cite{footnote}
\begin{equation}
-\frac{\hbar^2}{2m}\sum_i\frac{\partial^2 \psi}{\partial z_i^2}-\frac{\hbar^2}{2m}\left(\sum_{i}\frac{\partial }{\partial z_i}\right)^2\psi
+i L \frac{\hbar^2}{m}\sum_{i}\frac{\partial \psi}{\partial z_i}+g\sum_{i>j}\delta(z_i-z_j)\psi=\left({\cal E}-\frac{\hbar^2 L^2}{2m}\right)\psi,
\end{equation}
which should be supplemented with the boundary conditions at $z_i=\{0,2\pi R\}$:
\begin{equation}
\psi(z_i=0)=\psi(z_i=2\pi R); \qquad \frac{\partial \psi}{\partial z_i}\bigg|^{z_i=0^+}_{z_i=2\pi R^-}= \frac{c m}{\hbar^2} \psi(z_i=0). 
\end{equation}
Note that this equation for $L=0$ has been obtained and discussed in~\cite{volosniev2017}. Now we assume that all of the bosons 
are in one state, i.e., we look for the wave function of the form $\psi=\prod_i \Phi(z_i)$. To minimize the energy, the function $\Phi(z)$ satisfies
the following non-linear Schr{\"o}dinger equation
\begin{equation}
-\frac{\hbar^2}{m}\frac{\partial^2\Phi}{\partial z^2}+i\frac{\hbar^2 L}{m}\frac{\partial \Phi}{\partial z}
-i \frac{\hbar^2 (N-1) A}{m} \frac{\partial\Phi}{\partial z} + g(N-1)|\Phi|^2\Phi+c\delta(z)\Phi=\mu\Phi,
\end{equation}
where $A=-i\int \Phi(x)^*\frac{\partial}{\partial x}\Phi(x)\mathrm{d}x$ defines the momentum of a boson, and $\mu$ is the chemical potential.
We rewrite this equation as
\begin{equation}
-\frac{\partial^2\Phi}{\partial z^2}+i l \frac{\partial \Phi}{\partial z} + \tilde g(N-1)|\Phi|^2\Phi+\tilde c\delta(z)\Phi=\tilde\mu\Phi,
\end{equation}
where $\tilde\mu=\frac{m \mu}{\hbar^2}$, $\tilde g=\frac{gm}{\hbar^2}$, $\tilde c=\frac{cm}{\hbar^2}$ and the momentum of the impurity is $P\equiv \hbar l=\hbar(L-A(N-1))$~\cite{footnote0}.
This non-linear equation has an analytic solution for the soliton-like behavior~\cite{tsuzuki1971, ishikawa1980,hakim1997, kamenev2016}, which determines the properties 
of the polaron in our problem. Due to the symmetry of the problem it is enough to consider only $0<z<\pi R$, where
\begin{equation}
\Phi=\sqrt{\frac{\tilde \mu}{\tilde g(N-1)}}\left(1-\beta \mathrm{sech}^2\left[\sqrt{\frac{\tilde\mu\beta}{2}}(z+z_0)\right]\right)^{\frac{1}{2}}e^{i\phi(z)},
\end{equation}
with 
\begin{equation}
\phi(z)=\mathrm{arctan}\left(\frac{\sqrt{\frac{2 l^2}{\tilde \mu}\beta}}{\mathrm{exp}\left[\sqrt{2\tilde \mu\beta}(z+z_0)\right]-2\beta+1}\right).
\end{equation}
Here $\beta=1- l^2/(2\tilde \mu)$. The corresponding chemical potential in the thermodynamic limit is found from the normalization condition $\int \Phi^2=1$,
\begin{equation}
\tilde \mu=\gamma \rho^2 \frac{N-1}{N}\left(1-2\sqrt{2\beta} \frac{ (\mathrm{tanh}(d)-1)}{\sqrt{\gamma}N}\right),
\end{equation}
where $\rho=N/(2\pi R)$, $\gamma=\tilde g/\rho$, and $d=\sqrt{\frac{(2\gamma\rho^2-l^2)}{4}} z_0$.
The condition on $z_0$ is found by matching the boundary conditions at $z=\{0,2\pi R\}$
\begin{equation}
\frac{\tilde c}{\rho \sqrt{2\gamma}}=\frac{\beta^{\frac{3}{2}}\tanh(d)}{-\beta+\cosh^2(d)},
\end{equation}
where $\beta$ is taken in the limit $N\to\infty$.
The equation is cubic (in $\tanh(d)$), hence, the solutions can be found in a closed form;
there are two relevant solutions, which we refer to as the polaron and the soliton-polaron pair,
the latter is unstable (cf.~Ref.~\cite{hakim1997}) and we do not discuss it here.
Now we can calculate the energy of the polaron, which we define as
\begin{equation}
E=\lim_{N\to\infty, \frac{N}{2\pi R}\to \rho}\left[{\cal E}(c,P)-{\cal E}(c=0,P=0)\right],
\end{equation}
where
\begin{equation}
{\cal E}(c,P) =  \mu N-\frac{\hbar^2A^2N(N-1)}{2m}-gN(N-1)\int_{0}^{\pi R}|\Phi|^4\mathrm{d}z+\frac{\hbar^2 L^2}{2m}.
\end{equation}
Using these expressions we derive 
\begin{equation}
E=\frac{P^2}{2m}+\frac{\sqrt{2\gamma \beta}\rho^2\hbar^2 }{3 m}\left[4 b + (-4b+\beta\mathrm{sech}^2(d))\tanh(d)\right]+\lim_{N\to\infty}\frac{l\hbar^2A(N-1)}{m},
\end{equation}
where $b=1+\frac{l^2}{4\tilde g\rho}$, and $\beta$ is taken in the limit $N\to\infty$. This energy for $l\to 0$ can be written as 
\begin{equation}
E\simeq \frac{P^2}{2m}+\epsilon+\alpha P^2,
\end{equation}
where $\epsilon$ was derived in~Ref.~\cite{volosniev2017} (but not the effective mass). 
This equation allows us to introduce the effective mass as $1/m_{\mathrm{eff}}=1/m+2 \alpha$; 
see Fig.~1 of the main text.

\section{Quench dynamics from the effective Hamiltonian}
\label{app:b}

Here we discuss the dynamics of the impurity, which at $t=0$ is in the ground state of the initial Hamiltonian, i.e., 
\begin{equation}
\phi(y,t=0)=\left(\frac{m\Omega}{\pi\hbar}\right)^{\frac{1}{4}}e^{-\frac{m\Omega y^2}{2\hbar}},
\end{equation}
and at $t>0$ its evolution is determined by the Hamiltonian
\begin{equation}
\overline H^{k}_{\mathrm{eff}}=\overline \epsilon-\frac{\hbar^2}{2\overline m_{\mathrm{eff}}}\frac{\partial^2}{\partial y^2}+\frac{\overline m_{\mathrm{eff}} \Omega_{\mathrm{eff}}^2 y^2}{2}.
\end{equation}
For simplicity in this section we set $\overline \epsilon=0$ and $\hbar=m=1$. The wave function is written as
\begin{equation}
\phi(y,t)=\int\mathrm{d}y' K(y,y',t)\phi(y',0),
\end{equation}
where $K(y,y',t)$ is the propagator (see, e.g., Refs.~\cite{feynman1965, Barone2003})
\begin{equation}
K(y,y',t)=\left(\frac{\Omega_{\mathrm{eff}} \overline m_{\mathrm{eff}}}{2 \pi I \sin(\Omega_{\mathrm{eff}} t)}\right)^{\frac{1}{2}} e^{\frac{i\Omega_{\mathrm{eff}} \overline m_{\mathrm{eff}}}{2\sin(\Omega_{\mathrm{eff}} t)}\left[\left(y^2+y'^2\right)\cos(\Omega_{\mathrm{eff}} t)-2 yy'\right]}.
\end{equation}
The integrand is a Gaussian function, therefore, the integral can be easily computed. The wave function $\phi(y,t)$ is
\begin{equation}
\phi(y,t)=\left(\sqrt{\Omega}\frac{\Omega_{\mathrm{eff}} \overline m_{\mathrm{eff}}}{ i \sqrt{\pi}\sin(\Omega_{\mathrm{eff}} t)}\right)^{\frac{1}{2}}\frac{e^{\frac{\overline m_{\mathrm{eff}}y^2\Omega_{\mathrm{eff}}(-i\overline m_{\mathrm{eff}}\Omega_{\mathrm{eff}}-\Omega\cot(\Omega_{\mathrm{eff}}t))}{2 i \Omega+2 \overline m_{\mathrm{eff}} \Omega_{\mathrm{eff}}\cot(\Omega_{\mathrm{eff}}t)}}}{\sqrt{\Omega-i \overline m_{\mathrm{eff}}\Omega_{\mathrm{eff}} \cot(\Omega_{\mathrm{eff}}t)}}.
\end{equation}
This function can be used to calculate observables, e.g., the density.

\section{Many-Body Numerical Approach}
\label{app:c}

Let us briefly discuss the main features of our computational methodology: 
The Multi-Layer Multi-Configuration Time-Dependent Hartree Method for Atomic Mixtures (ML-MCTDHX) \cite{cao2013,mistakidis2017}. 
The ML-MCTDHX is a variational method for investigating stationary properties and nonequilibrium 
quantum dynamics of time-dependent many-body Schr{\"o}dinger equations that describe 
Bose-Bose \cite{Mistakidis_cor,Katsimiga_DBs}, Fermi-Fermi \cite{Koutentakis_FF} 
and Bose-Fermi mixtures \cite{expansion}. 
A key feature of this method is the expansion of the total many-body wavefunction in a 
time-dependent and variationally optimized basis, which enables one to efficiently track the system's important 
intra- and interspecies correlations with a computationally feasible basis size. 
In other words, it allows us to span the relevant (for the system under consideration) subspace 
of the Hilbert space at each time instant using a reduced number of basis states when 
compared to the expansions that rely on time-independent bases. 
Below, we elaborate on the many-body wavefunction ansatz used to simulate the nonequilibrium 
dynamics in the main text.

To account for interspecies correlations, we introduce $M$ distinct species functions for each component, 
i.e. $\Psi^{A}_k (\vec x;t)$ and $\Psi^{B}_k (y;t)$  where $\vec x=\left( x_1, \dots, x_{N} \right)$ and $y$  
refer to the spatial coordinates of the bosons (below denoted as $A$) and impurity (below denoted as $B$), respectively. 
Note that $\{\Psi_k^{\sigma}\}$ forms an orthonormal $N_{\sigma}$-body wavefunction set in a subspace of the $\sigma=A,B$ 
species Hilbert space $\mathcal{H}^{\sigma}$. 
The many-body wavefunction is then expressed according to the truncated Schmidt decomposition \cite{Horodecki} of rank $M$ 
\begin{equation}
\Psi_{MB}(\vec x, y;t) = \sum_{k=1}^M \sqrt{ \lambda_k(t) }~ \Psi^A_k (\vec x;t) \Psi^B_k (y;t),   
\label{Eq:WF}
\end{equation}
where the Schmidt coefficients $\lambda_k(t)$ in decreasing order are referred to as the natural species populations 
of the $k$-th species function. 
To infer about the presence of interspecies correlations or entanglement we employ the eigenvalues 
$\lambda_k$ of the species reduced density matrix which e.g. for the $N$ boson species reads
$\rho^{N} (\vec{x}, \vec{x}';t)=\int dy \Psi^*_{MB}(\vec{x},y;t) \Psi_{MB}(\vec{x}',y;t)$. 
The system is said to be entangled \cite{Roncaglia} or interspecies correlated when at least two distinct $\lambda_k(t)$ 
are macroscopically populated otherwise it is termed non-entangled. 
In the presence of entanglement the many-body state cannot be expressed as a direct product of two states and a certain 
configuration of $A$ species, $\Psi_k^A(\vec x;t)$, is accompanied by a particular configuration of $B$ 
species, $\Psi_k^B(y;t)$, and vice versa. 

Furthermore in order to incorporate intraspecies correlations each species function is expanded with respect to the permanents 
of $m^{A}$, $m^{B}$ different time-dependent single-particle functions ($\varphi_1,\dots,\varphi_{m^{A}}$), ($\varphi_1,\dots,\varphi_{m^{B}}$) respectively. 
Then, for the majority component consisting of $N$ bosons the expansion reads 
\begin{equation}
\Psi_k^{A}(\vec x;t) = \sum_{\substack{n_1,\dots,n_{m^{A}}\\
\sum n_i=N}} C^A_{k,(n_1,
\dots,n_{m^{A}})}(t) \sum_{i=1}^{N!} \mathcal{P}_i
 \left[ \prod_{j=1}^{n_1} \varphi_1(x_j;t) \cdots \prod_{j=1}^{n_{m^{A}}} \varphi_{m^{A}}(x_j;t) \right],   
 \label{Eq:SPFs1}
\end{equation} 
while for the impurity it becomes 
\begin{equation}
\Psi_k^{B}(y;t) = \sum_{l=1}^{m^B} C^B_{k,(n_1=0,
\dots,n_l=1,\dots,n_{m^{B}}=0)}(t) \varphi_l(y;t).  
 \label{Eq:SPFs2}
\end{equation} 
In Eq. (\ref{Eq:SPFs1}), $\mathcal{P}$ denotes the permutation operator that exchanges the particle configuration within 
the single-particle functions. 
Moreover, in both expressions $n_i$, $i=1,\dots,m^{\sigma}$ is the occupation number of the single-particle function $\varphi_i(\vec{x};t)$ 
for the $\sigma=A$ and $\varphi_i(y;t)$ for the $\sigma=B$ species, while 
$C^A_{k,(n_1,\dots,n_{m^{A}})}(t)$ and $C^B_{k,(n_1,\dots,n_{m^{B}})}(t)$ correspond to the time-dependent expansion coefficients of a particular 
permanent of the $A$ and $B$ component respectively.  
Utilizing a variational principle, e.g., the Dirac-Frenkel \cite{Frenkel,Dirac}, for the above-mentioned ansatz 
(see Eqs.~(\ref{Eq:WF}), (\ref{Eq:SPFs1}), and (\ref{Eq:SPFs2})) we obtain the corresponding ML-MCTDHX equations of 
motion \cite{cao2013,mistakidis2017}. 
The latter consist of a set of $M^2$ ordinary linear differential equations of motion for the coefficients $\lambda_k(t)$, 
coupled to a set of $M$[${N+m^A-1}\choose{m^A-1}$+${m^B}\choose{m^B-1}$] non-linear integro-differential equations for 
the species functions, and $m^A+m^B$ non-linear integro-differential equations for the single-particle functions.

Note that the ML-MCTDHX provides us with the opportunity to operate within different approximation schemes.
For instance, the commonly used mean-field scenario can be easily studied with the choice $M=m^A=m^B=1$. 
Indeed, for this choice the many-body wavefunction ansatz reduces to 
the mean-field product state  
\begin{equation}
\Psi_{MF} (\vec x,y;t) = \Psi^B_{MF} (y;t)\prod_{i}\Psi^A_{MF} (x_i;t). \label{Eq:MF}
\end{equation} 
Here, $\Psi_{MF}^{A}(x;t)$ and $\Psi_{MF}^{B}(y;t)$ are time-dependent functions.
Following the Dirac-Frenkel variational principle \cite{Dirac,Frenkel}, the mean-field ansatz~(\ref{Eq:MF}) leads to the celebrated 
set of the coupled Gross-Pitaevskii equations~\cite{pethick,stringari,emergent}  
\begin{equation}
\begin{split}
 i\hbar \frac{\partial \Psi_{MF}^A(x,t)}{\partial t}= \left[-\frac{\hbar^2}{2m}\frac{\partial^2}{\partial x^2}+\frac{kx^2}{2}+g\abs{\Psi_{MF}^A(x,t)}^2+c\abs{\Psi_{MF}^B(y,t)^2} \right] \Psi_{MF}^A(x,t),\\ 
 i\hbar \frac{\partial \Psi_{MF}^B(y,t)}{\partial t}= \left[-\frac{\hbar^2}{2m}\frac{\partial^2}{\partial y^2}+\frac{ky^2}{2}+c\abs{\Psi_{MF}^A(x,t)^2} \right] \Psi_{MF}^B(y,t).
\end{split}
\end{equation} 
Calculations based on this set of equations are presented in Fig.~2 of the main text. 
They disagree with the results of the ML-MCTDHB, which means that the beyond-mean-field 
correlations are important to describe accurately the dynamics of an impurity in a Bose gas.

Another commonly used approach is called the species mean-field approximation~\cite{cao2013,mistakidis2017,Katsimiga_DBs,Mistakidis_cor}, in which $M=1$, i.e., the boson-impurity entanglement is neglected, but the boson-boson correlations are taken into account. 
The corresponding ansatz for the wave function is written as
\begin{equation}
\Psi_{SMF}=\Psi^A_{SMF}(\vec x;t)\Psi^B_{SMF}(y;t),
\end{equation}
where $\Psi^A_{SMF}(\vec x;t)$ is of the form given in Eq.~(\ref{Eq:SPFs1}). Using 
the Dirac-Frenkel variational principle \cite{Dirac,Frenkel} we derive time-dependent equations for $\Psi^A_{SMF}$ and $\Psi^B_{SMF}(y;t)$ 
\begin{align}
\label{eq:species_mean_field}
\int \mathrm{d} \vec x\left(\Psi^A_{SMF}(\vec x;t)\right)^* \left[H-i\hbar\frac{\partial}{\partial t}\right]\Psi^A_{SMF}(\vec x;t) \Psi^B_{SMF}(y;t) =0,\\
\int \mathrm{d} y \left(\Psi^B_{SMF}(y;t)\right)^* \left[H-i\hbar\frac{\partial}{\partial t}\right]\Psi^A_{SMF}(\vec x;t) \Psi^B_{SMF}(y;t) =0.
\end{align}
These equations are beyond the Gross-Pitaevskii model, but less complicated than the ML-MCTDHB equations.
For the impurity dynamics the species mean-field approach is more accurate that the Gross-Pitaevskii model, since it 
allows for a better description of the density of the Bose gas. 
This is seen from Eq.~(\ref{eq:species_mean_field}) that describes the dynamics of the impurity 
\begin{equation}
 i\hbar \frac{\partial \Psi_{SMF}^B(y,t)}{\partial t}-f(t) \Psi_{SMF}^B = 
\left[-\frac{\hbar^2}{2m}\frac{\partial^2}{\partial y^2}+\frac{ky^2}{2}+c \rho_{SMF}(y) \right] \Psi_{SMF}^B(y,t),
\end{equation}
where $\rho_{SMF}$ is the density of bosons within the species mean-field approach. The function $f(t)$ reads
\begin{equation}
f(t)=\int \mathrm{d} \vec x\left(\Psi^A_{SMF}\right)^* \left[\sum_{i}\left(-\frac{\hbar^2}{2m}\frac{\partial^2}{\partial x_i^2}+\frac{kx_i^2}{2}\right)+g\sum_{i>j}\delta(x_i-x_j)-i\hbar\frac{\partial}{\partial t}\right]\Psi^A_{SMF},
\end{equation}
it is real, hence, it affects only the phase of the wave function.
Since  $\rho_{SMF}$ is exact for $c=0$ and $m^A\to \infty$, the species mean-field approach includes, in principle, all correlations 
for $c\to 0, \forall g$ for the considered time intervals, but must be corrected for larger values of $c$.

We use the species mean-field approximation 
to appreciate the importance of the entanglement between the impurity and the environment. To this end, we study the size of the impurity cloud $\langle y^2 \rangle(t)/\langle y^2 \rangle(0)$,
see Fig.~\ref{Fig:species_mean_field}. The species mean-field results are close 
to those of the ML-MCTDHB for $c/g=0.1$ (see Fig.~\ref{Fig:species_mean_field}~{\bf (a)}), which implies that for this interaction strength it is sufficient to allow for beyond-mean-field correlations only between the bosons.
In sharp contrast, the boson-impurity entanglement is crucial for the time dynamics at $c/g=0.8$, see Fig.~\ref{Fig:species_mean_field}~{\bf (b)}.

%%%%%%%%%%%%%%%%%%%%%%%%%%%%%%%%%%%%%%%%%%%%%

\begin{figure*}[ht]
\includegraphics[width=1.0\textwidth]{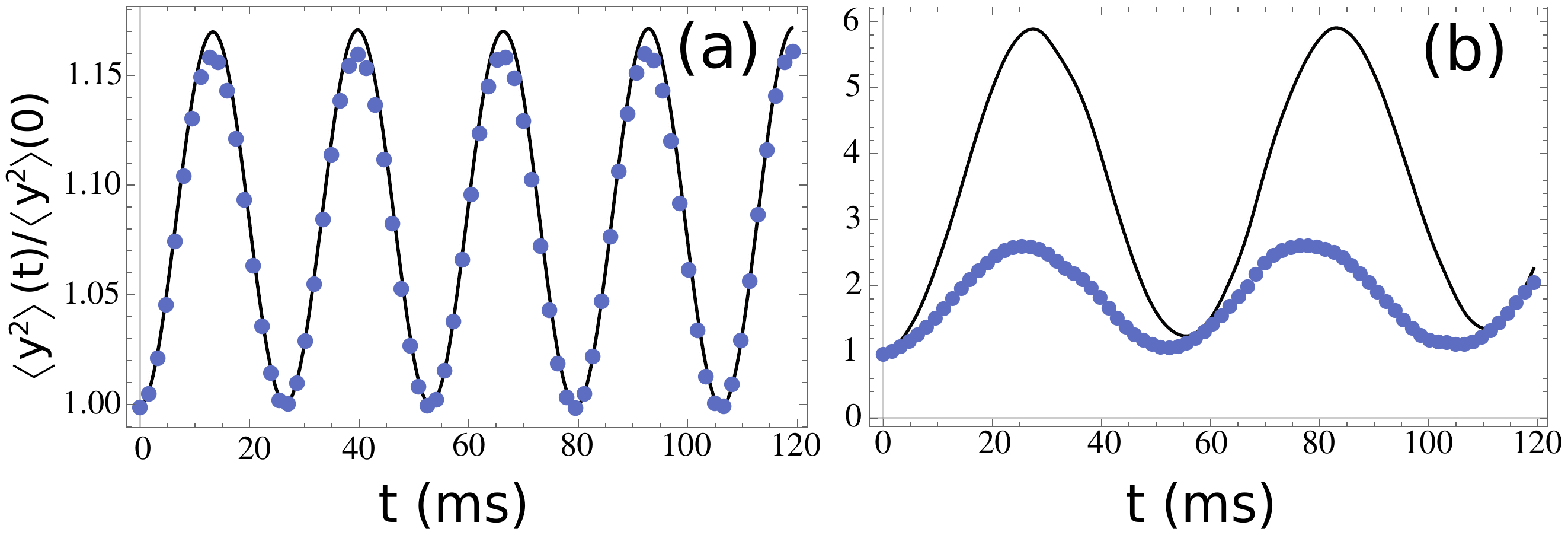}
\caption{The size of the impurity cloud $\langle y^2 \rangle(t)/\langle y^2 \rangle(0)$ as a function of time (in $m$s). 
The (black) solid curves show the ML-MCTDHB results. The (blue) dots present the corresponding species-mean-field approximation. 
Panel {\bf (a)} shows the results for $c/g=0.1$, and {\bf (b)} for $c/g=0.8$. All other parameters 
are as in the main text.
}
\label{Fig:species_mean_field}
\end{figure*}

%%%%%%%%%%%%%%%%%%%%%%%%%%%%%%%%%%%%%%%%%%%%%

%%%%%%%%%%%%%%%%%%%%%%%%%%%%%%%%%%%%%%%%%%%%%

\begin{figure*}[ht]
\includegraphics[width=1.0\textwidth]{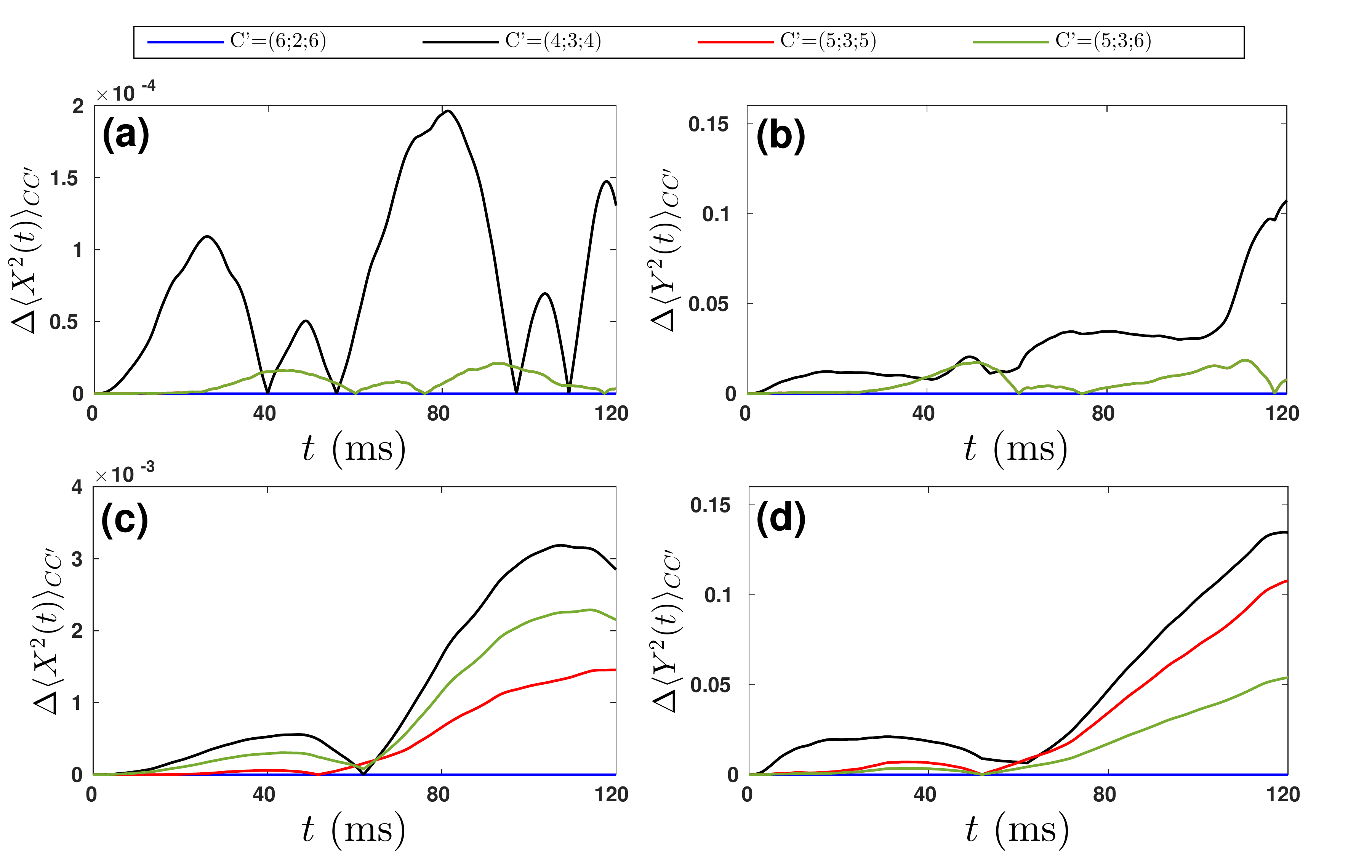}
\caption{Evolution of the relative difference of the variance $\Delta\braket{X^2(t)}_{CC'}$ and 
$\Delta\braket{Y^2(t)}_{CC'}$ for different orbital configurations 
$C=(6;3;6)$ and $C'$ (see legend) for the (a), (c) $A$ (bosons) and (b), (d) $B$ (impurity) species. 
In all cases the $A$-species consists of $N=100$ bosons. The boson-boson interaction strength is $g=10^{-37}$Jm. The figures
describe the quench dynamics upon the interaction change from $c=0$ to (a), (b) $c=0.8 g$ and (c), (d) $c=1.05 g$.}
\label{Fig:convergence}
\end{figure*}

%%%%%%%%%%%%%%%%%%%%%%%%%%%%%%%%%%%%%%%%%%%%%

\section{Ingredients and convergence of the Many-Body Simulations}
\label{app:d}

For our simulations, we employ a primitive basis consisting of a sine discrete variable representation 
that contains 450 grid points. 
To restrict the infinitely extended spatial region to a finite one, we impose hard-wall boundary conditions 
at the positions $x=\pm42\mu m$. 
This choice is justified by the fact that the Thomas-Fermi radius of the bosonic bath is of the order 
of 19$\mu m$ and we never observe significant density population beyond $x=\pm 27\mu m$. 
Consequently, the location of the boundary conditions does not affect our results. 
Moreover, the truncation of the total system's Hilbert space, i.e. the order of the considered approximation, 
is denoted by the used orbital configuration space $C=(M;m^A;m^B)$, where
$M$ is the number of species functions and $m^A$, $m^B$ are the corresponding numbers of single-particle 
functions used for each of the species. 
Regarding the numerical calculations presented in the main text we have used the configuration $C=(6;3;6)$, thus taking 
into acount 35028 coefficients of the available Hilbert space. 

Next, let us briefly elaborate on the convergence of our many-body calculations, and test the accuracy 
of the employed many-body truncation scheme. 
To this end, we vary the number of species functions and single-particle 
functions within the orbital configurations. This shows the degree of sensitivity of the considered observables 
to the employed approximation.
For the sake of brevity here we present the convergence of our main observable, namely the variance of each species, resorting to the 
relative difference of $\braket{X^2(t)}$, $\braket{Y^2(t)}$ between the different numerical configurations $C=(M;m^A;m^B)\equiv (6;3;6)$ 
and $C'=(M';m{^A}';m{^B}')$
\begin{equation}
\Delta\braket{X^2(t)}_{CC'}=\frac{\abs{\braket{X^2(t)}_{C}-\braket{X^2(t)}_{C'}}}{\braket{X^2(t)}_{C}} ~~\mathrm{and}~~ 
\Delta\braket{Y^2(t)}_{CC'}=\frac{\abs{\braket{Y^2(t)}_{C}-\braket{Y^2(t)}_{C'}}}{\braket{Y^2(t)}_{C}}.
\label{Eq:rel_dif} 
\end{equation} 
In Eq. (\ref{Eq:rel_dif}), $\braket{X^2(t)}=\bra{\Psi_{MB}(t)}\hat{x}^2\ket{\Psi_{MB}(t)}- \bra{\Psi_{MB}(t)}\hat{x}\ket{\Psi_{MB}(t)}^2$ and 
$\braket{Y^2(t)}=\bra{\Psi_{MB}(t)}\hat{y}^2\ket{\Psi_{MB}(t)}- \bra{\Psi_{MB}(t)}\hat{y}\ket{\Psi_{MB}(t)}^2$. 
Here, $\hat{x}=\int_{D}dx~x~\hat{\Psi}_{A}^\dagger(x)\hat{\Psi}_{A}(x)$, $\hat{y}=\int_{D}dy~y~\hat{\Psi}_{B}^\dagger(x)\hat{\Psi}_{B}(x)$ and 
$\hat{x}^{2}=\int_{D}dx~x^2~\hat{\Psi}_{A}^\dagger(x)\hat{\Psi}_{A}(x)$, $\hat{y}^{2}=\int_{D}dy~y^2~\hat{\Psi}_{B}^\dagger(x)\hat{\Psi}_{B}(x)$ 
are one-body operators, while $\hat{\Psi}_{A}(x)$ and $\hat{\Psi}_{B}(x)$ refer to the $A$ and $B$ species field operator respectively, and $D$ is the spatial extent of integration. 
The case of $\Delta\braket{X^2(t)}_{CC'}\to 0$, $\Delta\braket{Y^2(t)}_{CC'}\to 0$ indicates 
negligible deviations in $\braket{X^2(t)}$, $\braket{Y^2(t)}$ calculated within the $C$ and $C'$ approximations. 
Figure~\ref{Fig:convergence} shows $\Delta\braket{X^2(t)}_{CC'}$ and $\Delta\braket{Y^2(t)}_{CC'}$ for $N=100$ and the impurity atom with $g=1$ 
following an interspecies interaction quench from $c=0$ to $c=0.8 g$ ($c<g$) and $c=1.05g$ ($c>g$). 
Note that we fix $C=(6;3;6)$ and examine the convergence upon varying either $M'$ or $m^{A'}$, $m^{B'}$. 
We argue that our results at small times are accurate since both $\Delta\braket{X^2(t)}_{CC'}$ 
and $\Delta\braket{Y^2(t)}_{CC'}$ acquire relatively small values during the evolution.   
For the change to $c=0.8 g$ [Figs. \ref{Fig:convergence} (a), (b)] we observe that $\Delta\braket{x^2(t)}_{CC'}$ (bath component) 
exhibits negligible deviations being smaller than $1\%$ throughout the dynamics, while $\Delta\braket{y^2(t)}_{CC'}$ (impurity component) reaches a 
maximum value of the order of $9\%$ at large propagation times. 
The same observations hold also for $\Delta\braket{X^2(t)}_{CC'}$ and $\Delta\braket{Y^2(t)}_{CC'}$ for the change to $c=1.05 g$;
see Figs. \ref{Fig:convergence} (c), (d). 
Here, we observe that $\Delta\braket{X^2(t)}_{CC'}$ [$\Delta\braket{Y^2(t)}_{CC'}$] testifies relatively small deviations 
which become at most $1\%$ [$11\%$] at long evolution times. 
We should also note that similar results can be obtained upon testing other observables considered in the 
present effort e.g. the fidelity of the impurity (findings not shown here for brevity).

\section{Time evolution of the probability density of the impurity atom}
\label{app:e}

Here we discuss another observable, which can be easily determined using
the proposed effective Hamiltonian: the probability density for an impurity particle, $\rho_I(y)$.
First, we determine the ground state densities of the Hamiltonian $H$ for $c/g=0.1$ and $c/g=0.3$ using ML-MCTDHB
to show that those agree well with the predictions of the effective Hamiltonian, see Fig.~\ref{Fig:density_ground}.
Second, we investigate the densities during the quench dynamics discussed in the main text. We focus on the changes $c/g=0 \to c/g=0.1$ and $c/g=0 \to c/g=0.9$. 
For the former case, the densities obtained by ML-MCTDHB and the effective Hamiltonian 
agree well for all values of $t$ within the time interval [0, 120 $m$s]. The most noticeable deviation occurs around $y=0$, but even there the
relative difference is smaller than 1\% for all times. 
The case $c/g=0.9$ is more interesting, and we discuss it in more detail, see Fig.~\ref{Fig:density_t}.
At small values of $t$ the ML-MCTDHB results agree with the predictions of the effective Hamiltonian.
At $t=40 ms$ there is an overall agreement on the relevant scales (e.g., the size of the cloud), but the shapes of the densities are different. 
This is not surprising, since at $t=40 ms$ the impurity can probe the edges of the cloud, which is beyond the effective Hamiltonian.
At $t=80 ms$ the impurity is closer to the center of the trap, and the densities agree better than at $t=40 ms$.

%%%%%%%%%%%%%%%%%%%%%%%%%%%%%%%%%%%%%%%%%%%%%

\begin{figure*}[ht]
\includegraphics[width=0.5\textwidth]{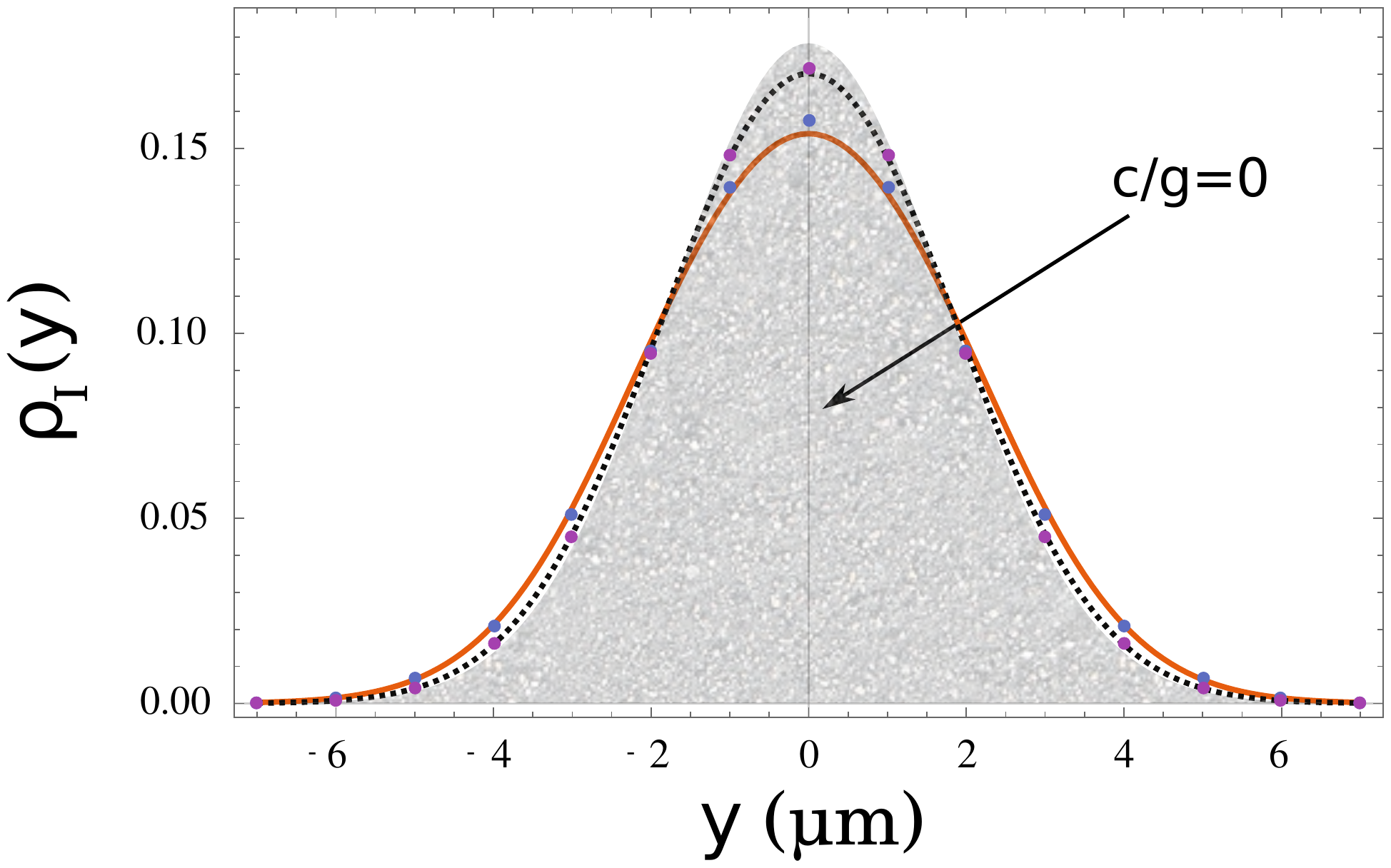}
\caption{The density of the impurity $\rho_I$ as a function of the coordinate $y$. The curves 
are calculated using ML-MCTDHB with the parameters extracted from the quench dynamics, see the main text. 
The curves are for $c/g=0.1$ (dotted, black) and $0.3$ (solid, orange). The dots are the corresponding densities 
from the effective Hamiltonian. For comparison, we also present the density of a non-interacting impurity ($c/g=0$)
as a grey shaded area.
}
\label{Fig:density_ground}
\end{figure*}

%%%%%%%%%%%%%%%%%%%%%%%%%%%%%%%%%%%%%%%%%%%%%

%%%%%%%%%%%%%%%%%%%%%%%%%%%%%%%%%%%%%%%%%%%%%

\begin{figure*}[ht]
\includegraphics[width=1\textwidth]{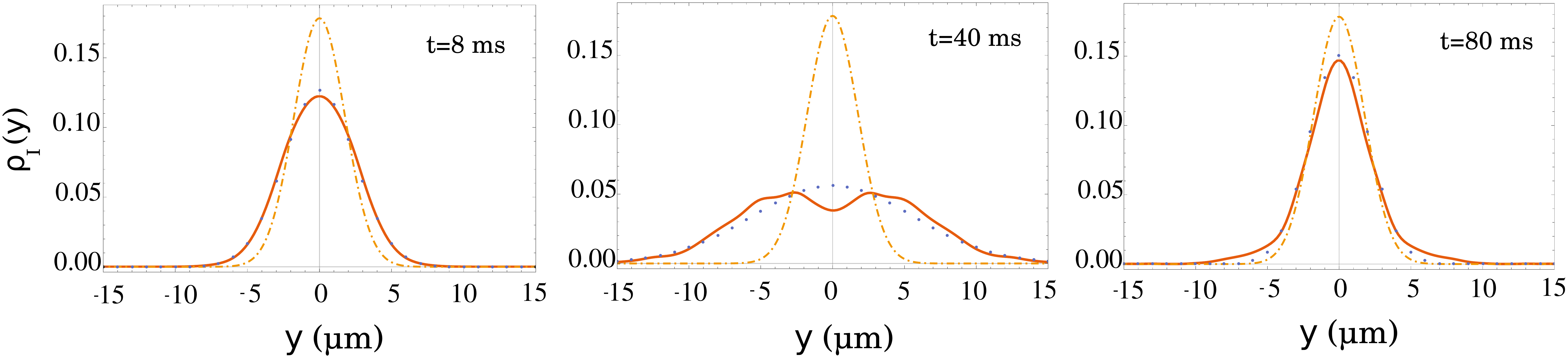}
\caption{The density of the impurity at $t=8,40$ and $80$ ms. The dynamics is initiated by a sudden change 
of the interaction strength: $c/g=0 \to c/g=0.9$. The solid curves present the ML-MCTDHB results. The dots are the results of the effective Hamiltonian
with the parameters according to the main text. For comparison, we also present the initial density of the impurity (i.e., at $t=0$)
as dot-dashed curves.}
\label{Fig:density_t}
\end{figure*}

%%%%%%%%%%%%%%%%%%%%%%%%%%%%%%%%%%%%%%%%%%%%%

\end{widetext}


\begin{thebibliography}{99}
\bibitem{pekar1951} S. Pekar. {\it Issledovanija po Elektronnoj Teorii Kristallov}. Gostekhizdat, Moskva (1951). English translation: {\it Research in Electron Theory of Crystals}, AEC-tr-555, US Atomic Energy Commission (1963).
\bibitem{polaron2000} N. N. Bogolubov and N. N. Bogolubov (Jr.), {\it Aspects Of Polaron Theory: Equilibrium and Nonequilibrium Problems}, World Scientific Publishing Company (2008). 
\bibitem{landau1948} L. D. Landau and S. I. Pekar, JETP {\bf 18}, 419 (1948).
\bibitem{baym2008} G. Baym and C. Pethick, Landau Fermi-Liquid Theory: Concepts and Applications, John Wiley and Sons (2008). 
\bibitem{bishop1973} R. F. Bishop, Annals of Physics {\bf 78}, 391 (1973).
\bibitem{kutchera1993} M. Kutschera and W. W{\'o}jcik, Phys. Rev. C {\bf 47}, 1077 (1993).
\bibitem{braun2017} A. Braun and Q. Chen, Nature Commun. {\bf 8}, 15830 (2017).
\bibitem{salamon2001} M. B. Salamon and M. Jaime, Rev. Mod. Phys. {\bf 73} , 583 (2001).
\bibitem{lee2006} P. A. Lee, N. Nagaosa, and X.-G. Wen, Rev. Mod. Phys. {\bf 78}, 17 (2006).
\bibitem{gershenson2006} M. E. Gershenson, V. Podzorov, and A. F. Morpurgo, Rev. Mod. Phys. {\bf 78}, 973 (2006).
\bibitem{zwierlein2009} A. Schirotzek, C.-H. Wu, A. Sommer, and M. W. Zwierlein, Phys. Rev. Lett. {\bf 102}, 230402 (2009).
\bibitem{nascimbene2009} S. Nascimb{\`e}ne, N. Navon, K. J. Jiang, L. Tarruell, M. Teichmann, J. McKeever, F. Chevy, and C. Salomon,
Phys. Rev. Lett. {\bf 103}, 170402  (2009).
\bibitem{chevy2010} F. Chevy and C. Mora, Rep. Prog. Phys. {\bf 73}, 112401 (2010).
\bibitem{widera2012} N. Spethmann, F. Kindermann, S. John, C. Weber, D. Meschede, and A. Widera, Phys. Rev. Lett. {\bf 109}, 235301 (2012). 
\bibitem{grimm2012} C. Kohstall, M. Zaccanti, M. Jag, A. Trenkwalder, P. Massignan, G. M. Bruun, F. Schreck, and R. Grimm, Nature {\bf 485}, 615 (2012).
\bibitem{koschorreck2012} M. Koschorreck, D. Pertot, E. Vogt, B. Fr{\"o}hlich, M. Feld, and  M. K{\"o}hl, Nature {\bf 485}, 619 (2012).
\bibitem{massignan2014} P. Massignan, M. Zaccanti, G. M. Bruun, Rep. Prog. Phys. {\bf 77}, 034401 (2014).
\bibitem{arlt2016} N. B. J{\o}rgensen, L. Wacker, K. T. Skalmstang, M. M. Parish, J. Levinsen, R. S. Christensen, G. M. Bruun, J. J. Arlt, Phys. Rev. Lett. {\bf 117}, 055302 (2016).
\bibitem{jin2016} M.-G. Hu, M. J. Van de Graaff, D. Kedar, J. P. Corson, E. A. Cornell, and D. S. Jin, Phys. Rev. Lett. {\bf 117}, 055301 (2016).
\bibitem{scazza2017} F. Scazza, G. Valtolina, P. Massignan, A. Recati, A. Amico, A. Burchianti, C. Fort, M. Inguscio, M. Zaccanti, and G. Roati, Phys. Rev. Lett. {\bf 118}, 083602 (2017).
\bibitem{schmidt2018}  R. Schmidt, M. Knap, D. A. Ivanov, J.-S. You, M. Cetina, and E. Demler, Rep. Prog. Phys. {\bf 81}, 024401 (2018). 
\bibitem{tajima2018} H. Tajima and S. Uchino, New J. Phys. {\bf 20}, 073048 (2018).



\bibitem{wenz2013} A. N. Wenz, G. Z{\"u}rn, S. Murmann, I. Brouzos, T. Lompe, S. Jochim, Science {\bf 342} 457 (2013).
\bibitem{astrakharchik2013} G. E. Astrakharchik and I. Brouzos, Phys. Rev. A {\bf 88}, 021602(R) (2013).
\bibitem{lindgren2014} E. J. Lindgren, J. Rotureau, C. Forss{\'e}n, A. G. Volosniev, N. T. Zinner, New J. of Phys. {\bf 16} 063003 (2014).
\bibitem{levinsen2015} J. Levinsen, P. Massignan, G. M. Bruun, and M. M. Parish, Science Advances {\bf 1}, e1500197 (2015).

\bibitem{catani2012} J. Catani, G. Lamporesi, D. Naik, M. Gring, M. Inguscio,
F. Minardi, A. Kantian, and T. Giamarchi, Phys. Rev. A {\bf 85}, 023623 (2012).


\bibitem{dehkharghani2015a} A.~S.~Dehkharghani, A.~G.~Volosniev, and N.~T.~Zinner, Phys. Rev. A {\bf 92}, 031601(R) (2015).
\bibitem{dehkharghani2018} A.~S.~Dehkharghani, A.~G.~Volosniev, and N.~T.~Zinner,  Phys. Rev. Lett. {\bf 121}, 080405 (2018). 
\bibitem{sartori2013} A. Sartori and A. Recati, Eur. Phys. J. D {\bf 67}, 260 (2013).
\bibitem{volosniev2015} A. G. Volosniev, H. W. Hammer, and N. T. Zinner, Phys. Rev. A {\bf 92}, 023623 (2015).
\bibitem{akram2016} J. Akram and A. Pelster, Phys. Rev. A {\bf 93}, 033610 (2016).
\bibitem{kamenev2016} M. Schecter, D. M. Gangardt, and A. Kamenev, New J. Phys. {\bf 18} 065002 (2016).
\bibitem{grusdt2017} F. Grusdt, G. E. Astrakharchik, and E. A. Demler, New J. Phys. {\bf 19}, 103035 (2017).
\bibitem{lampo2017} A. Lampo, S. Hoe Lim, M. {\'A}. Garc{\'i}a-March, and M. Lewenstein, Quantum {\bf 1}, 30 (2017). 
\bibitem{cao2013} L. Cao, S. Kr{\"o}nke, O. Vendrell, and P. Schmelcher, J. Chem. Phys. {\bf 139}, 134103 (2013).
\bibitem{mistakidis2017} L. Cao, V. Bolsinger, S. I. Mistakidis, G. M. Koutentakis, S. Kr{\"o}nke, J. M. Schurer, and P. Schmelcher,
J. Chem. Phys. {\bf 147}, 044106 (2017).
\bibitem{Mistakidis_cor} S.I. Mistakidis, G.C. Katsimiga, P.G. Kevrekidis, and P. Schmelcher, 
New J. Phys. \textbf{20}, 043052 (2018). 
\bibitem{Katsimiga_DBs} G.C. Katsimiga, G.M. Koutentakis, S.I. Mistakidis, P.G. Kevrekidis, and P. Schmelcher, 
New J. Phys. \textbf{19}, 073004 (2017). 
\bibitem{Koutentakis_FF} G.M. Koutentakis, S.I. Mistakidis, and P. Schmelcher, 
New J. Phys. {\bf 21}, 053005 (2019).  
\bibitem{expansion} P. Siegl, S.I. Mistakidis, and P. Schmelcher, 
Phys. Rev. A \textbf{97}, 053626 (2018).

\bibitem{kohl2009} S. Palzer, C. Zipkes, C. Sias, and M. K{\"o}hl, Phys. Rev.
Lett. {\bf 103}, 150601 (2009).

\bibitem{kuhr2013} T. Fukuhara, A. Kantian, M. Endres, M. Cheneau, P. Schaul{\ss}, S. Hild, D. Bellem, U. Schollw{\"o}ck, T. Giamarchi, C. Gross, I. Bloch, and S. Kuhr,
Nature Physics {\bf 9}, 235 (2013).
\bibitem{knap2017} F. Meinert, M. Knap, E. Kirilov, K. Jag-Lauber,
M. B. Zvonarev, E. Demler, H.-C. N{\"a}gerl, Science {\bf 356}, 945 (2017).


\bibitem{parisi2017} L. Parisi and S. Giorgini, Phys. Rev. A {\bf 95}, 023619 (2017).
\bibitem{volosniev2017} A. G. Volosniev and H.-W. Hammer, Phys. Rev. A {\bf 96}, 031601(R) (2017).  
\bibitem{pastukhov2017} V. Pastukhov, Phys. Rev. A {\bf 96}, 043625 (2017). 
\bibitem{robinson2016} N. J. Robinson, J.-S. Caux, and R. M. Konik,
Phys. Rev. Lett. {\bf 116}, 145302  (2016).
\bibitem{campbell2017} A. S. Campbell and D. M. Gangardt, SciPost Phys. {\bf 3}, 015 (2017).
\bibitem{astra2004} G. E. Astrakharchik and L. P. Pitaevskii, Phys. Rev. A
{\bf 70}, 013608 (2004).
\bibitem{sacha2006} K. Sacha, and E. Timmermans, Phys. Rev. A {\bf 73}, 063604 (2006).
\bibitem{bruderer2008} M. Bruderer, W. Bao, and D. Jaksch, EPL {\bf 82} 30004 (2008).
\bibitem{kain2016} B. Kain and H. Y. Ling, Phys. Rev. A {\bf 94}, 013621 (2016).
\bibitem{olshanii1998} M. Olshanii, Phys. Rev. Lett. {\bf 81}, 938 (1998).
\bibitem{chin2010} C. Chin, R. Grimm, P. Julienne, and E. Tiesinga, Rev. Mod. Phys. {\bf 82}, 1225 (2010).
\bibitem{tsuzuki1971} T. Tsuzuki, Journal of Low Temperature Physics 4, 441 (1971).
\bibitem{ishikawa1980} M. Ishikawa and H. Takayama, J. of Phys. Soc. of Jap. {\bf 49}, 1242 (1980).
\bibitem{hakim1997} V. Hakim, Phys. Rev. E {\bf 55}, 2835 (1997). 
\bibitem{pethick} C. J. Pethick and H. S. Smith, \emph{Bose-Einstein Condensation in Dilute Gases}, Cambridge University Press, Cambridge, (2002).
\bibitem{timmermans1998} E. Timmermans, Phys. Rev. Lett. {\bf 81}, 5718 (1998).
\bibitem{malomed2000} M. Trippenbach, K. G{\'o}ral, K. Rzazewski, B. Malomed, and Y. B. Band, J. Phys. B: At. Mol. Opt. Phys. {\bf 33}, 4017 (2000).
\bibitem{feynman1965} R. P. Feynman and A. R. Hibbs, {\it Quantum Mechanics and Path Integrals}. McGraw–Hill, New York (1965).
\bibitem{Barone2003} F. A. Barone, H. Boschi-Filho, and C. Farina, Am. J. of Phys. {\bf 71}, 483 (2003).
\bibitem{egorov2013} M. Egorov, B. Opanchuk, P. Drummond, B. V. Hall, P. Hannaford, and A. I. Sidorov, Phys. Rev. A {\bf 87}, 053614 (2013).
\bibitem{sykes2009} A. G. Sykes, M. J. Davis, and D. C. Roberts, Phys. Rev. Lett. {\bf 103}, 085302 (2009).
\bibitem{lausch2018} Tobias Lausch, Artur Widera, and Michael Fleischhauer, Phys. Rev. A {\bf 97}, 033620 (2018). 
\bibitem{peotta2013} Sebastiano Peotta, Davide Rossini, Marco Polini, Francesco Minardi, and Rosario Fazio, Phys. Rev. Lett. {\bf 110}, 015302 (2013).

\bibitem{mistakidis_20190} S. I. Mistakidis, G. C. Katsimiga, G. M. Koutentakis, Th. Busch, P. Schmelcher, Phys. Rev. Lett. {\bf 122},  183001 (2019).

\bibitem{kasevich2016} K. Sakmann and M. Kasevich, Nature Phys. {\bf 12}, 451 (2016).
\bibitem{marchukov2017} O. V. Marchukov and U. R. Fischer, Annals Phys. {\bf 405}, 274 (2019).

\bibitem{stringari} L. P. Pitaevskii and S. Stringari,
{\it Bose-Einstein Condensation}, Oxford University Press (Oxford, 2003).
\bibitem{emergent}
P. G. Kevrekidis, D. J. Frantzeskakis, and R. Carretero-Gonz{\'a}lez (eds.),
{\it Emergent nonlinear phenomena in Bose-Einstein condensates. Theory and experiment}
(Springer-Verlag, Berlin, 2008).
\bibitem{footnote1} The Hamiltonian must be corrected when the harmonic oscillator length associated with the trap~in Eq.~(\ref{eq:eff_k0}) is of the order of $R$, i.e., $\sqrt{\frac{\hbar}{m\Omega\sqrt{1-c/g}}}\simeq R$. For our system it leads to $c\simeq 0.99g$.  

\bibitem{landau5} L. D. Landau and E. M. Lifshitz, {\it Statistical Physics. Part 1.}, Pergamon Press (Oxford) (1980).
\bibitem{footnote_density} For $g\to 0$, this condition leads to the Thomas-Fermi profile since $E\simeq g N \rho/2$.
For $g\to\infty$, it leads to the density in the fermionized limit~\cite{girardeau1960} (cf.~Ref.~\cite{qi2000}) since $E=N\hbar^2\pi^2\rho^2/(6m)$.
\bibitem{kim2003} Yeong E. Kim and Alexander L. Zubarev, Phys. Rev. A {\bf 67}, 015602 (2003).
\bibitem{girardeau1960} M. D. Girardeau, J. Math. Phys. {\bf 1}, 516 (1960).
\bibitem{qi2000} E. B. Kolomeisky, T. J. Newman, J. P. Straley, and X. Qi, Phys. Rev. Lett. {\bf 85}, 1146 (2000). 
\bibitem{lieb1963} E. H. Lieb and W. Liniger, Phys. Rev. {\bf 130}, 1605 (1963).
\bibitem{schmidt2007} Bernd Schmidt and Michael Fleischhauer,  Phys. Rev. A {\bf 75}, 021601(R) (2007).
\bibitem{footnote_Bog} The correction using the Bogoliobov theory~\cite{parisi2017}: $-\sqrt{\frac{m\rho(x)}{\pi^2\hbar^2 g}}c^2$,  agrees well with our result. 
\bibitem{fetter1998} A. Fetter and D. Feder, Phys. Rev. A {\bf 58}, 3185 (1998). 
\bibitem{smith2019} D. H. Smith and A. G. Volosniev, {\bf arXiv}:1903.11462 (2019).
\bibitem{panochko2019} G. Panochko and V. Pastukhov, {\bf arXiv}:1903.05953 (2019).
\bibitem{Horodecki} R. Horodecki, P. Horodecki, M. Horodecki, and K. Horodecki,  
%Quantum entanglement. 
Rev. Mod. Phys. \textbf{81}, 865 (2009). 


\bibitem{Roncaglia} M. Roncaglia, A. Montorsi, and M. Genovese, Phys. Rev. A \textbf{90}, 062303 (2014). 


\bibitem{Frenkel} J. Frenkel, {\it in Wave Mechanics} 1st ed. (Clarendon Press, Oxford, 1934), pp. 423-428.

\bibitem{Dirac} P. A. Dirac,  
Proc. Camb. Phil. Soc., 
\textbf{26}, 376, Cambridge University Press (1930).
\bibitem{footnote} The conservation of $L$ allows 
us to reduce the polaron problem to an $N$-body problem that describes correlations between particles.
The position of a particle itself contains no information about the polaron due to the rotational symmetry of a ring.
The reduction due to the conservation of $L$ is very similar to the decoupling of the center-of-mass coordinate in infinite homogeneous spaces. 

\bibitem{footnote0} 
The momentum of the impurity is actually $\hbar(L-A N)$, but this difference is not important in the thermodynamic limit, which is of our interest.



\end{thebibliography}
\end{document}